\documentclass[a4paper]{aa}  
\usepackage{graphicx}
\usepackage{natbib}
\usepackage{txfonts}
\def\SiII{\mbox{\ion{Si}{ii}}}
\def\ForbOI{\mbox{[\ion{O}{i}]}}

\def\Ha{\ifmmode^{\mathrm{H}\alpha }\else$\mathrm{H}\alpha$\fi}
\def\Hb{\ifmmode^{\mathrm{H}\beta }\else$\mathrm{H}\beta$\fi}
\def\LyA{\ifmmode^{\mathrm{H}\alpha }\else$\mathrm{Ly}\alpha$\fi}
\def\BrA{\ifmmode^{\mathrm{Br}\alpha }\else$\mathrm{Br}\alpha$\fi}
\def\BrG{\ifmmode^{\mathrm{Br}\gamma }\else$\mathrm{Br}\gamma$\fi}
\def\PaB{\ifmmode^{\mathrm{Pa}\beta }\else$\mathrm{Pa}\beta$\fi}
\def\mag{\ifmmode^{\rm m }\else$^{\rm m}$\fi}

\def\as{$\,^{\prime\prime}\,$}

\def\hh{\ifmmode^{\rm h}\else$^{\rm h}$\fi}
\def\mm{\ifmmode^{\rm m}\else$^{\rm m}$\fi}
\def\ss{\ifmmode^{\rm s}\else$^{\rm s}$\fi}
\def\deg{\ifmmode^\circ\else$^\circ $\fi}
\def\amin{\ifmmode^\prime\else$^\prime $\fi}

\def\decdm#1#2{\ifmmode{#1}\else{$#1$}\fi\deg\ #2\amin\ }

\def\dec#1#2#3{\ifmmode{#1}\else{$#1$}\fi\deg\ #2\amin\ #3\as\ }

\def\decb#1#2#3#4{\ifmmode{#1}\else{$#1$}\fi\deg\ #2\amin\ #3\farcs#4 }

\begin{document}

\title{
  The origin of hydrogen line emission for five Herbig~Ae/Be stars spatially
  resolved by VLTI/AMBER spectro-interferometry\thanks{Based on observations
    made with ESO telescopes at the La Silla Paranal Observatory under open
    time programme IDs 077.C-0694, 078.C-0360, and 078.C-0680. }
}

\titlerunning{Origin of hydrogen line emission in Herbig~Ae/Be stars}

\author{
  S.~Kraus\inst{1} \and
  K.-H.~Hofmann\inst{1} \and
  M.~Benisty\inst{2} \and
  J.-P.~Berger\inst{2} \and
  O.~Chesneau\inst{3} \and
  A.~Isella\inst{4} \and
  F.~Malbet\inst{2} \and
  A.~Meilland\inst{1} \and
  N.~Nardetto\inst{1} \and
  A.~Natta\inst{5} \and
  T.~Preibisch\inst{1} \and
  D.~Schertl\inst{1} \and
  M.~Smith\inst{6} \and
  P.~Stee\inst{3} \and
  E.~Tatulli\inst{2} \and
  L.~Testi\inst{7} \and
  G.~Weigelt\inst{1}
}

\authorrunning{Kraus~et~al.}

\offprints{skraus@mpifr-bonn.mpg.de}

\institute{
  Max Planck Institut f\"ur Radioastronomie, Auf dem H\"ugel 69, 53121 Bonn, Germany\\
  \email{skraus@mpifr-bonn.mpg.de} \and
  Laboratoire d'Astrophysique de Grenoble, UMR 5571 Universit\'{e} Joseph Fourier/CNRS, BP 53, 38041 Grenoble Cedex 9, France\and
  UMR 6525 H. Fizeau, Univ.\ Nice Sophia Antipolis, CNRS, Observatoire de la C\^{o}te d'Azur, Av.\ Copernic, F-06130 Grasse, France\and
  Caltech, MC 105-24, 1200 East California Blvd., Pasadena CA 91125, USA\and
  INAF-Osservatorio Astrofisico di Arcetri, Largo Fermi 5, 50125 Firenze, Italy\and
  Centre for Astrophysics \& Planetary Science, University of Kent, Canterbury CT2 7NH, UK\and
  European Southern Observatory, Karl-Schwarzschild-Strasse 2, 85748 Garching, Germany
}

\date{Received 2008-04-10; accepted 2008-07-03}

\abstract
{
  Accretion and outflow processes are of fundamental importance for our
  understanding of the formation of stars and planetary systems.
  To trace these processes, diagnostic spectral lines such as the {\BrG} $2.166~\mu$m
  line are widely used, although due to a lack of spatial resolution, 
  the origin of the line emission is still unclear.
}
{
  Employing the AU-scale spatial resolution which can be achieved with infrared
  long-baseline interferometry, we aim to distinguish between theoretical
  models which associate the {\BrG} line emission with mass infall
  (magnetospheric accretion, gaseous inner disks) or mass outflow processes
  (stellar winds, X-winds, or disk winds).
}
{
  Using the VLTI/AMBER instrument, we spatially and spectrally
  ($\lambda/\Delta\lambda=1500$) resolved the inner ($\lesssim 5$~AU) environment 
  of five Herbig~Ae/Be stars (HD\,163296, HD\,104237, HD\,98922, MWC\,297, V921\,Sco)
  in the {\BrG} emission line as well as in the adjacent continuum. 
  From the measured wavelength-dependent visibilities, we derive the
  characteristic size of the continuum and {\BrG} line-emitting region.
  Additional information is provided by the closure phase, which
  we could measure both in the continuum wavelength regime (for four objects)
  as well as in the spectrally resolved {\BrG} emission line (for one object).
  The spectro-interferometric data is supplemented by archival and new VLT/ISAAC
  spectroscopy.
}
{
  For all objects (except MWC\,297), we measure
  an increase of visibility within the {\BrG} emission line, indicating that
  the {\BrG}-emitting region in these objects is more compact than the dust
  sublimation radius.  For HD\,98922, our quantitative analysis reveals that
  the line-emitting region is compact enough to be consistent with the magnetospheric accretion scenario.
  For HD\,163296, HD\,104237, MWC\,297, and V921\,Sco we identify an extended stellar
  wind or a disk wind as the most likely line-emitting mechanism.
  Since the stars in our sample cover a wide range of stellar parameters, we
  also search for general trends and find that the size of the
  {\BrG}-emitting region does not seem to depend on the basic stellar parameters (such 
  as the stellar luminosity), but correlates with spectroscopic properties, in particular
  with the {\Ha} line profile shape.
}
{
  By performing the first high-resolution spectro-interferometric survey on
  Herbig~Ae/Be stars, we find evidence for at least two distinct {\BrG} 
  line-formation mechanisms.
  Most significant, stars with a P-Cygni {\Ha}~line profile and a high
  mass-accretion rate seem to show particularly compact {\BrG}-emitting
  regions ($R_{\mathrm{Br}\gamma}/R_{\mathrm{cont}} <0.2$), while stars with
  a double-peaked or single-peaked {\Ha}-line profile show a
  significantly more extended {\BrG}-emitting region
  ($0.6 \lesssim R_{\mathrm{Br}\gamma}/R_{\mathrm{cont}} \lesssim 1.4$),
  possibly tracing a stellar wind or a disk wind.
}


\maketitle

%

\section{Introduction}
\label{sec:intro}

Accretion disks around young stellar objects (YSOs) are at the focus of 
astronomical research, not only because they play an essential role in
the star-formation process, but also because they provide the stage where planet
formation takes place.

Historically, these disks were discovered due to their characteristic infrared
excess emission, which is believed to trace dust grains, providing (for the
expected temperatures and densities) the dominant source of opacity.
In recent years, there has been substantial progress in constraining the
detailed three-dimensional geometry of the dust disk using, for example,
combined modeling of the spectral energy distribution (SED) and spatially
resolved infrared interferometric observations.
While the thermal emission from the dust disk is likely to be the dominant 
contributor to the infrared excess emission observed towards YSOs, it is
believed that the dust content makes up only a small fraction of the total mass
of the system. 99\% of the mass is likely contributed by gas, in
particular hydrogen, and can mainly be traced by the spectral lines, 
which for pre-main-sequence stars are often found in emission.  While some
information about the kinematics of the gas can already be extracted from the line
profile, the spatial origin of the gas emission and the physical processes
they trace are still strongly debated.

In this context, the Brackett-$\gamma$ ({\BrG}) 2.1661~$\mu$m line is of special importance.
It was found that the luminosity $L($\BrG$)$ of this line (as determined from the
circumstellar component of the line equivalent width) seems to correlate with the mass
accretion luminosity $L_{\mathrm{acc}}$, as determined from UV veiling.  
This empirical $L($\BrG$)$--$L_{\mathrm{acc}}$ relationship
has been established for pre-main-sequence stars of various masses, including brown dwarfs \citep{nat04}, 
T~Tauri~stars \citep{muz98a}, as well as intermediate-mass Herbig~Ae/Be stars \citep{cal04,van05}.
Based on this empirical correlation, observers started to use the {\BrG} 
luminosity as an estimator for the mass accretion rate \citep[e.g.][]{gar06},
which follows from $L_{\mathrm{acc}}$ using $\dot{M}_{\mathrm{acc}} = L_{\mathrm{acc}} R_{\star} / G M_{\star}$.

Therefore, it is now essential to identify the process(es) involved in the
formation of the {\BrG} emission line in YSOs.
The main scenarios which have been proposed up to now 
include mass infall, as well as mass outflow mechanisms:

\begin{figure*}[htbp]
  \centering
  \includegraphics[angle=0,width=18cm]{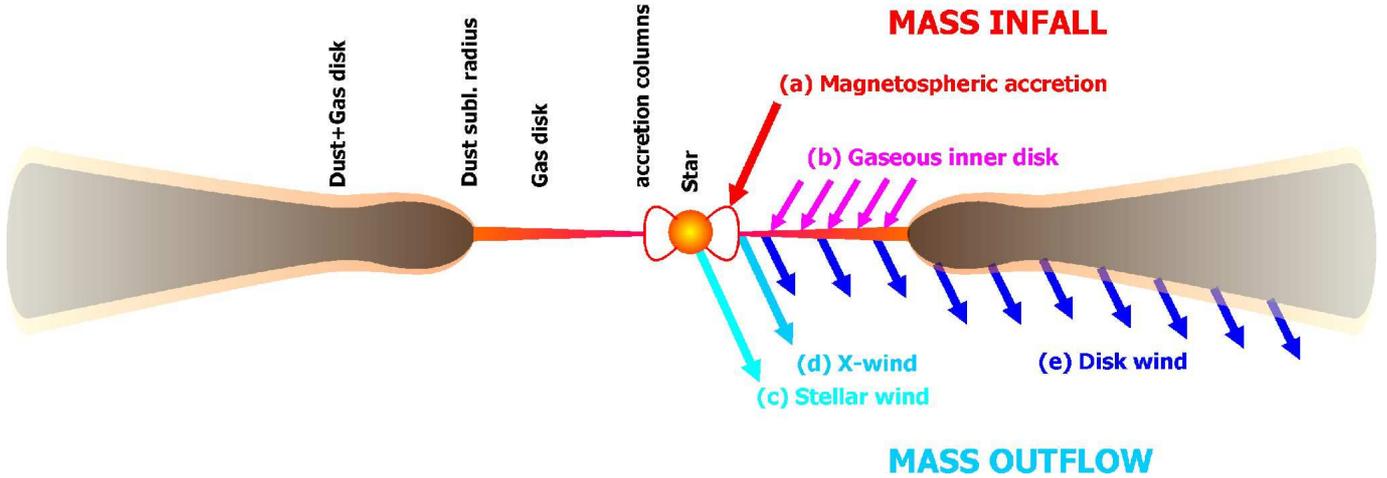}\\
  \caption{Illustration of the regions which have been proposed as the 
    origin of the permitted hydrogen recombination line emission observed
    towards HAeBe stars (this sketch is not to scale; read
    Sect.~\ref{sec:intro} for details about the individual mechanisms).}
  \label{fig:windorigin}
\end{figure*}

\noindent
{\bf (a)~Magnetospheric accretion:}
  The line emission might emerge from matter which is accreted onto the star
  through magnetospheric accretion columns \citep{van05}.  This infall is
  supposed to happen very close to the star, inside the co-rotation radius,
  where the Keplerian angular velocity matches the stellar angular velocity.

\noindent
{\bf (b)~Gaseous inner disk:}
  Inside of the dust destruction radius, the gas continues to accrete towards
  the star, forming a gaseous inner accretion disk.  The recombination line
  emission from ionized hydrogen in this disk might contribute to the line
  emission observed towards HAeBe stars.
  \citet{muz04} estimated that especially for high mass-accretion rates
  $\dot{M}_{\mathrm{acc}} > 10^{-7}~M_{\sun} $yr$^{-1}$, the flux contribution of
  the inner gaseous disk might be of importance.

\noindent
{\bf (c)~Stellar wind:}
  The P~Cygni line profile observed in the hydrogen lines of several
  HAeBe~stars might also indicate mass loss through stellar winds
  \citep[e.g.][]{mes68,cat87,str98}.
  While magnetically accelerated stellar winds seem plausible for massive
  stars rotating close to their break-up velocity, this scenario does not seem to
  work for lower-mass YSOs \citep{fer06}.

\noindent
{\bf (d)~Stellar-field driven wind (X-wind):}
  In the X-wind model \citep{shu94}, the outflows from YSOs are launched
  from a narrow interaction region where the stellar magnetosphere truncates
  the accretion disk.  Keplerian rotation leads to a
  winding up of the field lines and to the formation of magnetic
  surfaces. Charged particles from the stellar wind will then get trapped in
  these surfaces, accelerated, and collimated into a beam.

\noindent
{\bf (e)~Disk-field driven wind (disk wind):}
  Another mechanism which has been proposed for the launching and collimation
  of outflows and jets observed towards YSOs are magnetocentrifugally driven
  disk winds \citep{bla82,pud83,fer97}.
  Material from the disk surface of the rotating disk is centrifugally
  accelerated along the open magnetic field lines of the accretion disk.
  In contrast to the X-wind model, where the wind is launched from a narrow
  disk annulus around the inner truncation radius of the accretion disk, 
  disk winds originate from a wide range of radii, likely extending from the
  co-rotation radius out to several AUs.
\newline

Since most of these processes are believed to take place on (sub-)AU scales
and were therefore not accessible with direct imaging techniques, most earlier
studies tried to constrain the spatial distribution and kinematics of the
line-emitting gas from the shape of line profiles.  
However, since these line profile fitting techniques are known to be highly
ambiguous \citep[e.g.][]{cat99}, spatially resolved observations are urgently 
required to constrain the physical mechanism and spatial origin of important
tracer lines like {\BrG}.
Since the different emitting mechanisms noted
above can be associated to distinct spatial regions (see
Fig.~\ref{fig:windorigin}), spatially resolved observations should allow us to
identify and physically characterize the true underlying emitting mechanism.

The VLTI/AMBER instrument combines, for the first time, the milli-arcsecond
spatial resolution achievable with IR long-baseline interferometry  
with a good spectroscopic resolution ($R=\lambda/\Delta\lambda=1500$ or 12\,000)
in the near-infrared $K$-band. In this study, we use the unique capabilities
of this instrument in order to measure the geometry and position of the {\BrG}
line-emitting region relative to the continuum-emitting region, providing
direct information about the processes involved.

A particularly well-studied class of YSOs are the Herbig Ae/Be stars (HAeBes).
These are intermediate-mass, pre-main sequence stars.  Over the last few
years, the continuum emission from a rather large number of HAeBes could
already be studied with broad-band infrared interferometry.
These studies revealed a correlation between the size of
the NIR continuum emitting region and the stellar luminosity \citep{mon02},
suggesting that for most Herbig~Ae stars, the NIR emission likely traces
dust at the dust sublimation radius.

Only very little is known about the spatial origin of the hydrogen
recombination line emission.
First AMBER observations of the Herbig~Be star MWC\,297 have shown that 
the {\BrG}-emitting region around this star is more extended than the
continuum region \citep{mal07}.
Surprisingly, observations on the less luminous Herbig~Ae star
HD\,104237 did not show any change in visibility along the {\BrG} line
\citep{tat07a}. 
\citet{eis07b} found a visibility increase within the {\BrG} line for the
Herbig~Ae star MWC\,480.
This might suggest that the {\BrG} line traces fundamentally
different mechanisms for Herbig~Ae and Be stars. In this study, we present
observations on three HAeBes, which are resolved for the first time with
spectro-interferometry (HD\,163296, HD\,98922, V921\,Sco). For HD\,104237, we
present new observations which were obtained at longer baseline lengths,
adding substantial new information to the initial AMBER results presented by
\citet{tat07a}.  Finally, we re-reduced the archival data on MWC\,297 and
HD\,104237 in order to provide a uniformly calibrated sample of five stars for
our interpretation.  The stellar parameters which we adopted for
the stars in our sample are shown in Tab.~\ref{tab:targetstars}.

This paper is structured as follows:  In Sect.~\ref{sec:observations} we
discuss the spectro-interferometric and spectroscopic data and the applied
data reduction procedures, followed by an outline of the obtained results
(Sect.~\ref{sec:results}).
Then, we present the models which were fitted to the data in order to
constrain the geometry of the continuum- (Sect.~\ref{sec:modelingcont}) and
{\BrG}-emitting region (Sect.~\ref{sec:modelingbrg}).
After interpreting the individual objects in detail 
(Sect.~\ref{sec:interpretation}), we discuss general trends which we find in
our small sample of HAeBe stars (Sect.~\ref{sec:discussion}).
Finally, we conclude with a summary of our results and discuss the 
potential of future observations in Sect.~\ref{sec:conclusions}.

\section{Observations and data reduction}
\label{sec:observations}

\begin{table*}[t]
\caption{Target stars and adopted stellar parameters.}
\label{tab:targetstars}
\centering

\begin{tabular}{lccccccccc ccc}
  \hline\hline
  Star                & Spectral  & $L_{\star}$  & $d$    & $T_{\star}$   & $A_{V}$ & $R_{\star}$~$^{(a)}$ & $\log \frac{L(\mathrm{Br}\gamma)}{L_{\sun}}$~$^{(b)}$ & $\log \dot{M}^{\mathrm{Br}\gamma}_{\mathrm{acc}}$~$^{(c)}$ & $v \sin i$ & H$\alpha$~$^{(d)}$ & Ref.\\
                      & Type      & [$L_{\sun}$]  & [pc]        & [K]       &        & [$R_{\sun}$] &          & [$M_{\sun}$\,yr$^{-1}$]    &  [km\,s$^{-1}$] & profile & \\
  \noalign{\smallskip}
  \hline
  \noalign{\smallskip}
  \object{HD\,104237} & A5         & 30        & 116~$^{(f)}$   & 8\,000   & 0.31   & 2.9  & $-2.74$    & -7.45 & $12 \pm 2$~$^{(g)}$      & D$^{(h)}$ & $(i)$\\
  \object{HD\,163296} & A3         & 26        & 122~$^{(f)}$   & 8\,700   & 0.12   & 2.2  & $-2.78$    & -7.12 & $120^{+20}_{-30}$~$^{(j)}$ & D$^{(h)}$ & $(k)$\\
  \object{HD\,98922}  & B9         & 890       & 540~$^{(e)}$   & 10\,600  & 0.3    & 9.1  & $-1.56$    & -5.76 & --                       & P$^{(h)}$ & $(l)$\\
  \object{MWC\,297}   & B1.5       & 10600     & $250$         & 23\,700  & 8      & 6.1  & $>-0.6$    & --    & $350 \pm 50$~$^{(m)}$     & S$^{(h)}$ & $(m)$\\
  \object{V921\,Sco}  & B0         & 19950     & 800           & 30\,900  & 5.0    & 5.0  & $>-0.7$    & --    & --                       & D$^{(n)}$ & $(n)$\\
  \noalign{\smallskip}
  \hline
\end{tabular}

\begin{flushleft}
  \hspace{5mm}Notes~--~$^{(a)}$ The stellar radius $R_{\star}$ is computed using the
  given effective temperature and bolometric luminosity.\\ 
  \hspace{5mm}$^{(b)}$ Luminosity of the {\BrG} emission line, as determined by \citet{gar06} from ISAAC spectra.\\
  \hspace{5mm}$^{(c)}$ Mass accretion rate, as determined from the $L($\BrG$)$--$L_{\mathrm{acc}}$ relation \citep{gar06}.\\
  \hspace{5mm}$^{(d)}$ This column described the H$\alpha$-line profile shape of the stars in our sample using the classifications: D: double-peaked profile; P: P-Cygni profile; S: single-peaked profile.\\
  \hspace{5mm}$^{(e)}$ For HD\,98922, we assume the minimum distance given by \citet{gar06}.\\
  \hspace{5mm}References: $(f)$~\citealt{van98}; $(g)$~\citealt{don97}; $(h)$~\citealt{ack05}; $(i)$~\citealt{tat07a}; $(j)$~\citealt{fin85}; $(k)$~\citealt{van00a}; $(l)$~\citealt{gar06}; $(m)$~\citealt{dre97}; $(n)$~\citealt{hab03}\\ 
\end{flushleft}
\end{table*}

\subsection{VLT/ISAAC spectroscopy}
\label{sec:obsisaac}

The profile of the emission lines carries important
information about the kinematics of the emitting gas.  Therefore, we complement
the spatially resolved AMBER spectro-interferometry with high-spectral
resolution ($R\,\sim\,9000$) spectra obtained with the VLT/ISAAC instrument. 

Besides archival ISAAC data (ESO programme 073.C-0184, P.I.~E.~Habart), we
also obtained new spectroscopic data (for HD\,163296 and V921\,Sco, ESO
programme 077.C-0694, P.I.~S.~Kraus) in order to measure the line
profile as close in time to the AMBER observations as possible.  This new
spectroscopic data covers not only the {\BrG}~2.1661~$\mu$m line
(Fig.~\ref{fig:spectra}, left), but also the {\PaB}~1.2822~$\mu$m line
(Fig.~\ref{fig:spectra}, right).

The raw spectra were extracted using IRAF procedures and then corrected for
atmospheric features using telluric standard star observations obtained 
during the same night.
For wavelength-calibration, we aligned the raw spectra to publicly available 
high-resolution ($R=40000$) telluric spectra taken at the NSO/Kitt Peak 
Observatory.
To compare the final ISAAC spectra (Fig.~\ref{fig:spectra}) with the spectra extracted
from the AMBER data, we convolved the ISAAC spectra to the spectral
resolution of AMBER ($R=1500$) and found good agreement (see
Figs.~\ref{fig:hd104237}{\it(a)} to \ref{fig:v921sco}{\it(a)}, {\it top panel}).

\subsection{Archival ISO and {\it Spitzer}/IRS spectroscopy}
\label{sec:obsspectro}

In order to optimally constrain the SED for the stars in our sample, we
obtained near- to mid-infrared spectroscopic data from the ISO and 
{\it Spitzer} Space Telescope archive.
The {\it Spitzer}/IRS spectra were pre-processed by the S13.2.0 pipeline
version at the Spitzer Science Center (SSC) and then extracted with the SMART
software, Version 6.2.5 \citep{hig04}.

\subsection{VLTI/AMBER spectro-interferometry}
\label{sec:obsamber}

\begin{table*}[t]
  \caption{Observation log of our VLTI/AMBER interferometric observations with a spectral resolution $R=1500$.}
  \label{tab:observations}
  \centering

\begin{tabular}{lccccc|cc|cc|cc|lc}
  \hline\hline
  Target Star       & Date        & Time & Spectral        & DIT  & Telescope   & \multicolumn{6}{c|}{Projected Baselines} & Calibrator & Ref. \\
                    & (UT)        & (UT) & Window          & [ms] & Triplet     & $B_L$     & PA         & $B_L$ & PA        & $B_L$  & PA         &   &      \\
                    &             &      & [$\mu$m]        &      &             & [m]       & [$^\circ$] & [m]   & [$^\circ$] & [m]    & [$^\circ$] &   &      \\
  \noalign{\smallskip}
  \hline
  \noalign{\smallskip}
  HD\,104237       & 2005-02-26 & 07:14      & 2.12--2.20  & 100   & UT2-UT3-UT4    & 35.1 &  71  & (\,60.8 & 120\,) & (\,87.9 & 102\,)  & HD\,135382 & $(a)$\\
  HD\,104237       & 2007-01-09 & 08:03      & 1.94--2.26  & 200   & UT1-UT3-UT4    & (\,78.0 &  31\,)  & 58.5 & 84 & (\,122.4& 53\,)   & HD\,118934 & \\
  HD\,104237       & 2007-01-09 & 07:41      & 1.94--2.26  & 500   & UT1-UT3-UT4    & (\,78.7 &  26\,)  & 58.0 & 78 & (\,123.1& 47\,)   & HD\,118934 & \\
  HD\,163296       & 2006-04-13 & 06:19      & 2.12--2.20  & 50    & UT2-UT3-UT4    & 43.9 &  19  & 52.3 & 99 & 74.1 & 63   & HD\,171960 & \\
                   &            &            &             &       &                &      &      &      &    &      &      & HD\,89682 & \\
  HD\,98922        & 2007-02-04 & 07:19      & 2.12--2.20  & 50    & UT2-UT3-UT4    & 42.9 &  46  & (\,61.9 & 111\,)& (\,88.9 & 85\,)   & HD\,101328 & \\
  HD\,98922        & 2007-01-09 & 09:31      & 1.94--2.26  & 200   & UT1-UT3-UT4    & (\,90.5 &  42\,)  & 62.3 & 116& (\,122.8& 71\,)   & HD\,118934 & \\
  HD\,98922        & 2007-01-09 & 09:39      & 1.94--2.26  & 500   & UT1-UT3-UT4    & (\,89.9 &  43\,)  & 62.4 & 118& (\,122.0& 73\,)   & HD\,118934 & \\
  MWC\,297         & 2004-05-31 & 06:02      & 2.00--2.23  & 107   & UT2-UT3        & 44.7 &  42  & --   & -- & --   & --   & HD\,177756 & $(b)$\\
  V921\,Sco        & 2006-04-14 & 08:09      & 2.12--2.20  & 50    & UT2-UT3-UT4    & 45.2 &  42  & 62.1 & 110& (\,89.3 & 82\,)   & HD\,159941 & \\
  \noalign{\smallskip}
  \hline
\end{tabular}

\begin{flushleft}
  \hspace{5mm}Notes~--~For more detailed information about the calibrator stars, we refer to Tab.~\ref{tab:calibrators}.\\
  \hspace{5mm}References~--~$(a)$ reprocessing of data presented in \citet{tat07a}; $(b)$ reprocessing of data presented in \citet{mal07}.
\end{flushleft}
\end{table*}

\begin{table}[t]
\caption{Calibrator star information for the interferometric observations presented in Tab.~\ref{tab:observations}.}
\label{tab:calibrators}
\centering

\begin{tabular}{lccccc}
  \hline\hline
  Star                & $V$  & $K$  & Spectral & $d_{\mathrm{UD}}$\\
                      &      &      & Type     & [mas]           \\
  \noalign{\smallskip}
  \hline
  \noalign{\smallskip}
  \object{HD\,101328} & 7.44 & 3.75 & K4III    & $1.00 \pm 0.01$~$^{(a)}$\\
  \object{HD\,118934} & 7.92 & 4.02 & K4III    & $0.89 \pm 0.01$~$^{(a)}$\\
  \object{HD\,135382} & 2.88 & 2.53 & A1V      & $1.10 \pm 0.08$~$^{(b)}$\\
  \object{HD\,159941} & 7.85 & 3.55 & M0III    & $1.09 \pm 0.02$~$^{(a)}$\\
  \object{HD\,171960} & 7.29 & 3.38 & K3II     & $1.13 \pm 0.02$~$^{(a)}$\\
  \object{HD\,177756} & 3.43 & 3.56 & B9Vn     & $0.60 \pm 0.06$~$^{(b)}$\\
  \noalign{\smallskip}
  \hline
\end{tabular}

\begin{flushleft}
  \hspace{5mm}Notes~--~The $V$-band magnitudes were taken from SIMBAD and the $K$-band magnitudes from the 2MASS point source catalog.\\
  \hspace{5mm}References~--~$(a)$ UD diameter taken from the CHARM2
  catalog~\citep{ric05}. $(b)$ UD diameter computed with ASPRO.\\
\end{flushleft}
\end{table}

AMBER \citep{pet07} is the NIR beam-combiner of the Very Large Telescope
Interferometer (VLTI), which is located on Cerro Paranal/Chile and operated by
the European Southern Observatory (ESO). 
Combining the light from up to three of the four 8.4~m unit telescopes
simultaneously, AMBER measures not only visibility amplitudes, but
also the closure phase (CP) relation.  In the course of three ESO open time
programmes (077.C-0694, 078.C-0360, 078.C-0680, P.I.~S.~Kraus), we obtained
spectrally dispersed interferograms in AMBER's medium resolution (MR) mode
($R=1500$) on four HAeBes; namely, HD\,163296, HD\,104237,
HD\,98922, and V921\,Sco.  These new data sets were complemented with archival
data from MWC\,297 and HD\,104237, obtained earlier during AMBER commissioning and in
guaranteed-time observations \citep{mal07,tat07a}.  Re-reducing these data sets
with the latest software allows us to ensure homogenity both in data
reduction as well as in the applied modeling procedures. 
The observations are summarized in Tab.~\ref{tab:observations} and were 
obtained under good atmospheric conditions (seeing 0.5--0.9\arcsec,
atmospheric coherence time 2--7~ms). In order to 
calibrate the obtained visibilities and CPs for instrumental and
atmospheric effects, we used interferometric calibrator stars, taking the
intrinsic diameters of these stars into account (see
Tab.~\ref{tab:calibrators}).  By using different detector integration times
(DITs) for the different objects, we aimed for a compromise 
between collecting a sufficient number of photons and minimizing the loss of
fringe contrast due to atmospheric piston.  In general, observations with
longer DIT should provide better SNR in the differential visibility
measurement, while observations with shorter DIT should provide a better
absolute calibration.  Since we aimed mainly for a precise measurement of the
differential quantities in the {\BrG} and the adjacent continuum, we used 
rather long DIT for most observations.  In order to account for the resulting
errors on the absolute calibration, we estimate a rather large calibration
error of 5\%.
The spectral calibration of the data was done by comparing the spectrum
extracted from the AMBER data with ISAAC spectra of the same object (see
Sect.~\ref{sec:obsisaac}), which have been convolved to the spectral resolution
of AMBER.
An additional very important calibration step in the course of AMBER data
reduction was the relative spectral re-shifting of the photometric channels
with respect to the interferometric channel.  The appropriate shift was
determined to sub-pixel accuracy by computing the auto-correlation between
the spectra extracted from these channels.

In a first data reduction step, the AMBER raw interferograms were cleared from
the correlated detector noise effect using the AMDC software tool
\citep[Version~1.1,][]{cau07}. 
Then, the raw data was reduced with the \textit{amdlib2} software\footnote{The
  amdlib2 software package is available from the website\\ {\tt 
    http://www.jmmc.fr/data\_processing\_amber.htm}} (release 2.1), employing the P2VM
algorithm~\citep{tat07b}.
Due to the absence of a fringe tracker, a large fraction of the interferograms
is of rather low contrast (see discussion in \citealt{pet07}).  Therefore, we
removed any frames from our dataset for which the light injection from the
contributing telescopes was unsatisfying; i.e., the intensity ratio between
the photometric channels was larger than 4, or the fringe contrast was
decreased due to instrumental jitter (the 20\% best interferograms were
selected based on the Fringe SNR criteria, as defined in \citealt{tat07b}).

Since some stars in our sample are close to the current sensitivity limit of
AMBER's MR mode, we applied spectral binning to the raw data in order to
increase the fringe SNR.  
For each data set, the width of the sliding window was chosen so that the
resulting fringe SNR of the interferograms exceeds the critical value of 1.5,
ensuring a reliable visibility estimation.  Since the {\BrG}~line
for all our objects is spectrally resolved over several spectral channels at $R=1500$, the
resulting decrease in spectral resolution only marginally reduces the contrast between the
spectral line and the underlying continuum.
This procedure increases the fringe SNR significantly, while the loss in
spectral resolution (from $R=1500$ to 500 or 250) results only in a minor
decrease of the line-to-continuum flux ratio $F_{\mathrm{Br}\gamma}/F_{\mathrm{cont}}$.
Baselines for which a spectral binning to $R=250$ was not sufficient to yield
an SNR of 1.5 were rejected from further analysis (in Tab.~\ref{tab:observations}, 
these baselines are put in brackets).  The final visibility curves are shown in 
Figs.~\ref{fig:hd104237}{\it(a)} to \ref{fig:v921sco}{\it(a)} in the
second panel from the top. 

In principle, two distinct phase quantities can be extracted from AMBER interferograms, namely
differential phases (DP, measuring the relative displacement of the line-emitting region
with respect to the continuum-emitting region) and the closure phase, which can indicate
asymmetries in the source brightness distribution.
Since for most of our datasets, the fringe SNR in individual frames is not sufficient
to reliably correct for disturbing atmospheric phase contributions (atmospheric piston), 
no scientifically meaningful differential phases could be extracted, unfortunately.
However, for some data sets we could extract useful CP information, both in the 
continuum (HD\,104237, HD\,163296, HD\,98922, V921\,Sco) as well as in the line regime (V921\,Sco).
It is expected that the continuum CP does not change significantly over the 
small spectral window covered by our observations, which allowed us to apply an optimized data 
reduction strategy for the extraction of the continuum CPs.
First, we apply some spectral binning to the raw data (bracketing out the
spectral channels containing the emission line), followed by an averaging of the complex 
visibilities over various spectral channels (between 2.12--2.15 and 2.175--2.185~$\mu$m).
The obtained CPs will be presented in Sect.~\ref{sec:closurephases}.

\section{Results}
\label{sec:results}

\subsection{SED and NIR-spectroscopy}
\label{sec:SED}

\begin{figure}[htbp]
  \centering
  \includegraphics[angle=0,width=8.5cm]{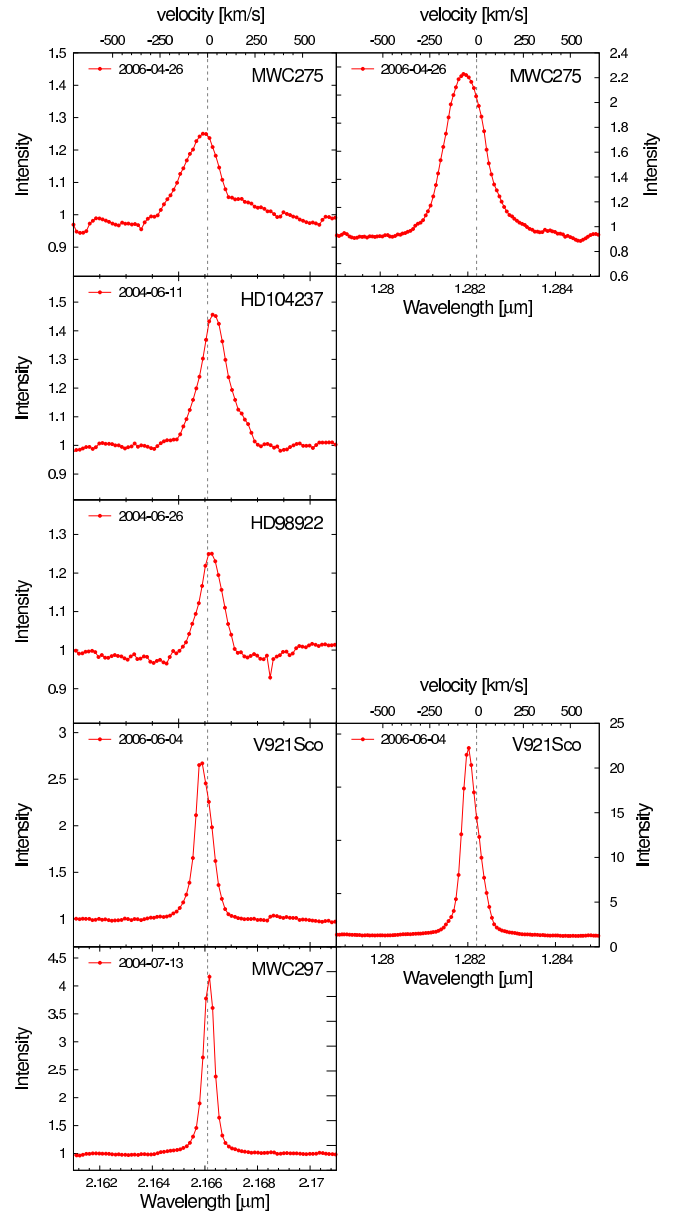}\\
  \caption{{\it Left:} VLT/ISAAC spectra showing the {\BrG} line with a spectral
    resolution of $R\sim9000$ for the stars in our sample.  
    {\it Right: } VLT/ISAAC spectra of the {\PaB} line were obtained for
    HD\,163296 and V921\,Sco.
  } 
  \label{fig:spectra}
\end{figure}

In order to constrain the SED of the stars in our sample, we collected
photometric and spectroscopic data from the literature and the ISO and {\it
  Spitzer} archives (see Sect.~\ref{sec:obsspectro}).
Assuming that the line-of-sight extinction can be mainly accounted to the
large scale envelopes, in which some HAeBe stars are embedded, we dereddened
the photometric and spectroscopic data assuming the values for $A_{V}$ given
in Tab.~\ref{tab:targetstars}, $R_{V}=3.1$, and the extinction law by
\citet{mat90}.
The derived SEDs (Figs.~\ref{fig:hd104237}{\it(b)} to \ref{fig:v921sco}{\it(b)})
allow us to determine the ratio between the star flux contribution
$f_{\rm star}$ and the flux contribution from the circumstellar disk $f_{\rm
  disk}$ for the fitting of geometric models (Sect.~\ref{sec:modelingcont}).

The VLT/ISAAC {\BrG} line spectra are shown in Fig.~\ref{fig:spectra}.
With the currently available resolution, all {\BrG} lines show a single-peaked
profile.  For HD\,163296 and V921\,Sco, measurements at two or three epochs are
available, indicating that the line profile of V921\,Sco has not 
significantly changed, while for HD\,163296 we find some variability.

\subsection{Spectro-interferometric visibilities}

\begin{figure*}[hp]
  \centering
  \includegraphics[angle=0,width=14cm]{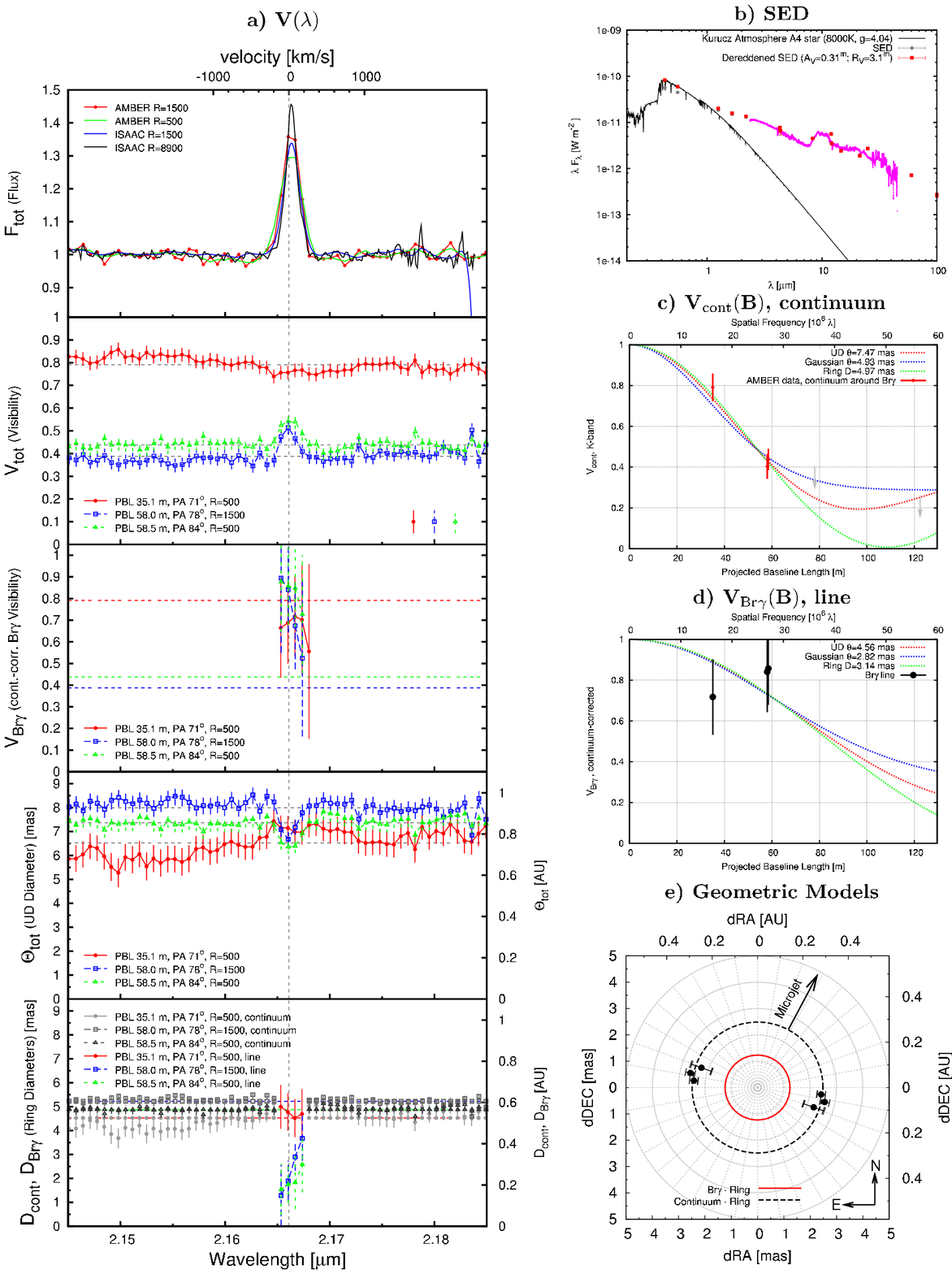}\\
  \caption{Spectroscopic, spectro-interferometric, and photometric data for
    HD\,104237: In {\it (a)}, we show the {\BrG} spectra extracted from our
    ISAAC and AMBER data {\it (top panel)}, the measured wavelength-dependent
    visibilities $V_{\mathrm{tot}}$ for different baselines (2nd panel, the data
    points are shown with statistical errors bars, whereas the estimated
    calibration errors are shown in the bottom-right corner), the
    continuum-corrected {\BrG}-line visibility $V_{\mathrm{Br}\gamma}$ (3rd panel), 
    the UD diameter $\Theta_{\mathrm{tot}}$ derived from $V_{\mathrm{tot}}$ (4th panel), 
    and the diameters $D_{\mathrm{cont}}$ and $D_{\mathrm{Br}\gamma}$ derived for the
    continuum- and the {\BrG}-emitting region using a 2-ring model (5th panel).
    In {\it (b)}, the SED is shown, including photometric data from
    the literature (grey points), archival ISO spectra (magneta), and
    archival {\it Spitzer}/IRS spectra (blue).
    In {\it (c)}, we plot the measured $K$-band continuum-visibility as a
    function of baseline length and in {\it (d)} the continuum-corrected
    {\BrG}-line visibility. Panel {\it (e)} shows the derived
      ring diameters for the continuum- and line-emitting region 
      plotted as function of position angle.}
  \label{fig:hd104237}
\end{figure*}

\begin{figure*}[h]
  \centering
  \includegraphics[angle=0,width=14cm]{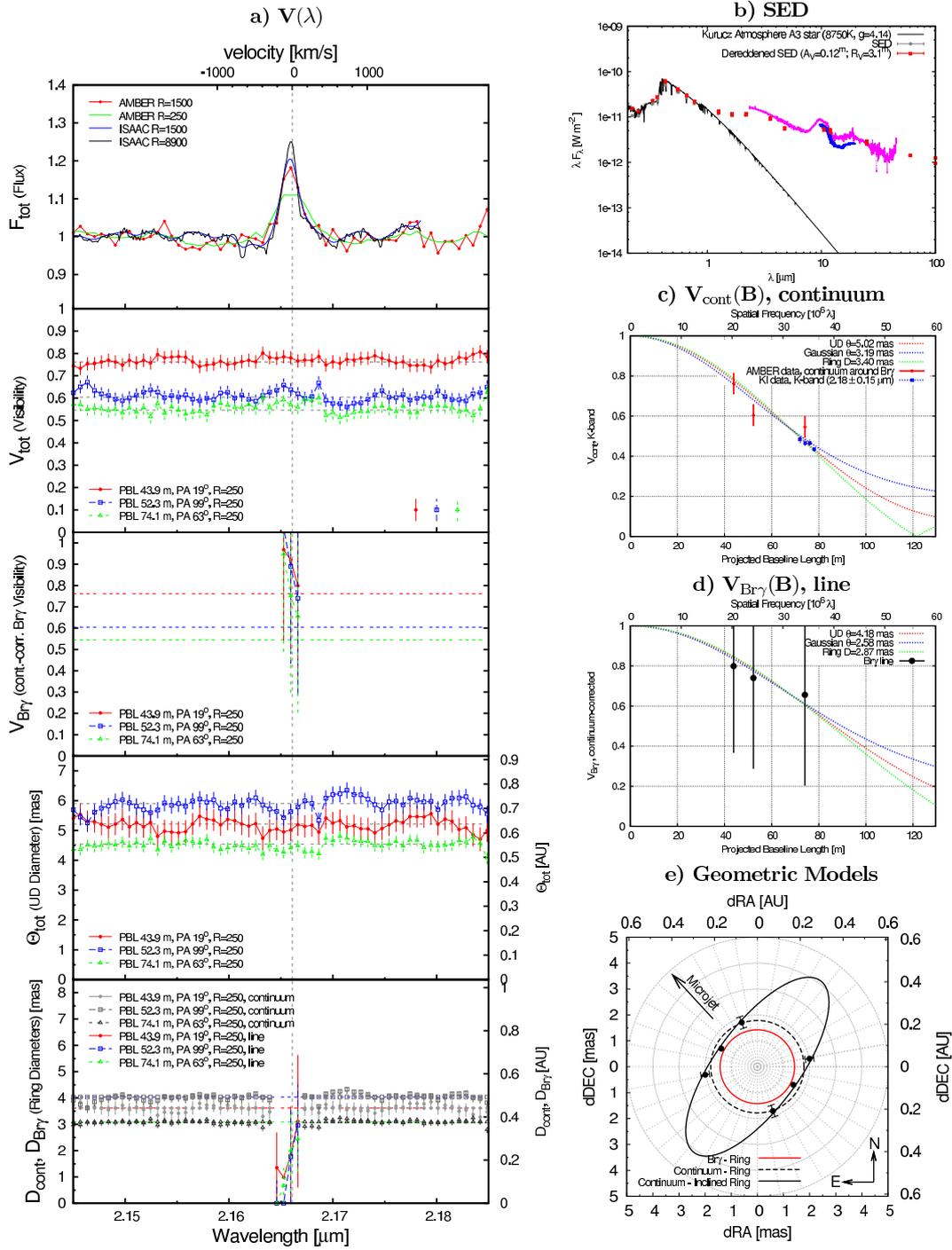}\\
  \caption{Spectroscopic, spectro-interferometric, and photometric data for
    HD\,163296 (similar to Fig.~\ref{fig:hd104237}).}
  \label{fig:mwc275}
\end{figure*}

\begin{figure*}[h]
  \centering
  \includegraphics[angle=0,width=14cm]{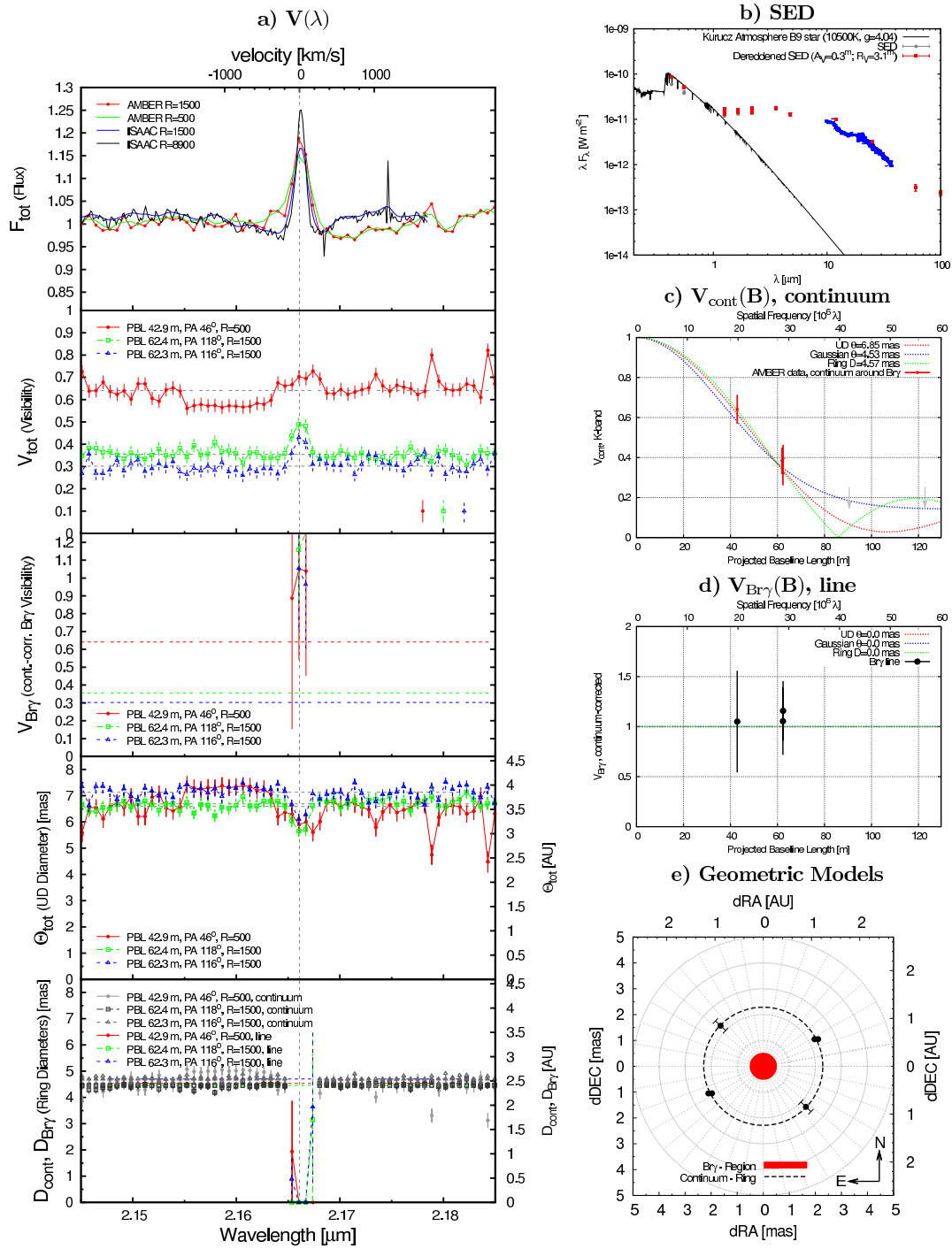}\\
  \caption{Spectroscopic, spectro-interferometric, and photometric data for
    HD\,98922 (similar to Fig.~\ref{fig:hd104237}). }
  \label{fig:hd98922}
\end{figure*}

\begin{figure*}[h]
  \centering
  \includegraphics[angle=0,width=14cm]{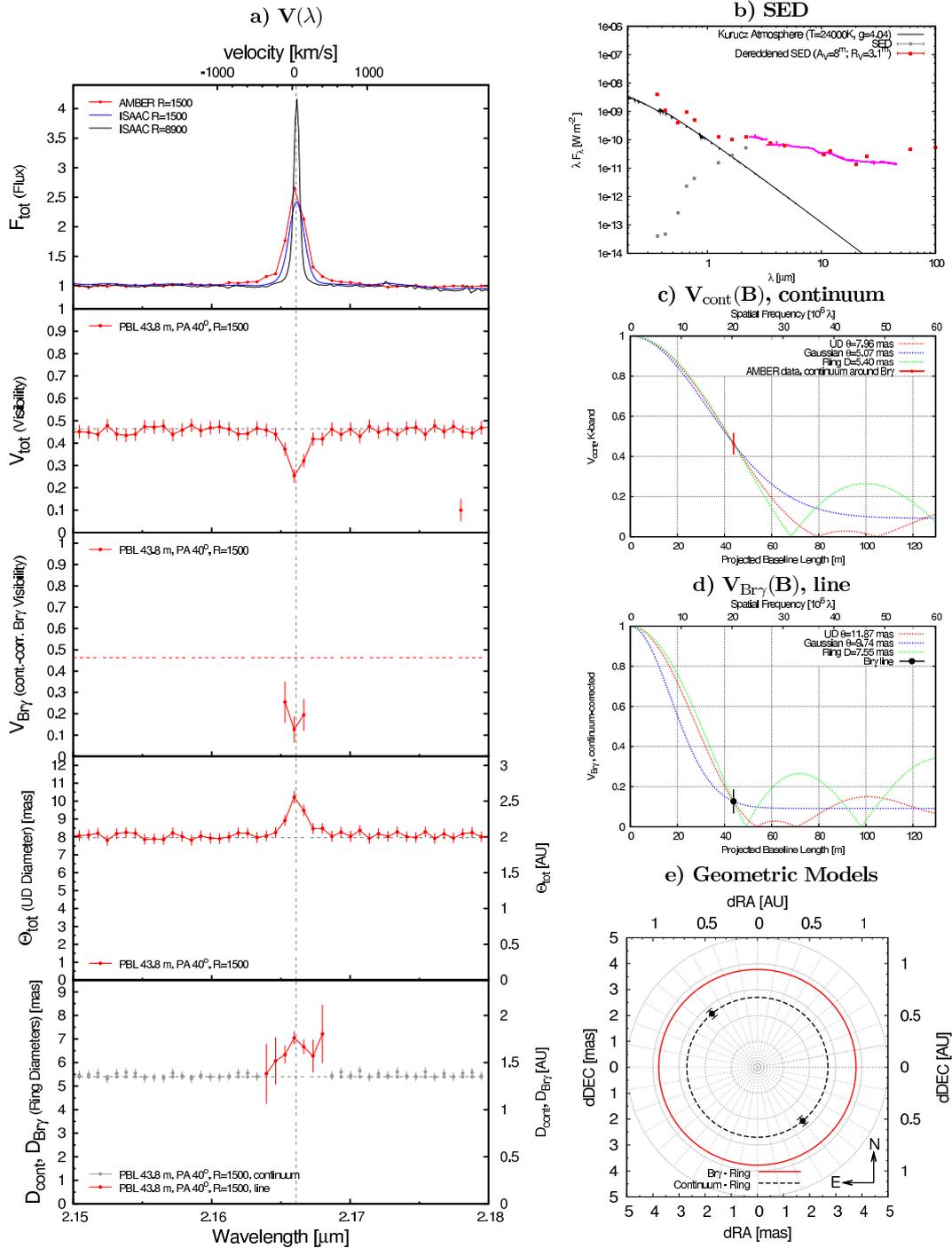}\\
  \caption{Spectroscopic, spectro-interferometric, and photometric data for
    MWC\,297 (similar to Fig.~\ref{fig:hd104237}). }
  \label{fig:mwc297}
\end{figure*}

\begin{figure*}[h]
  \centering
  \includegraphics[angle=0,width=14cm]{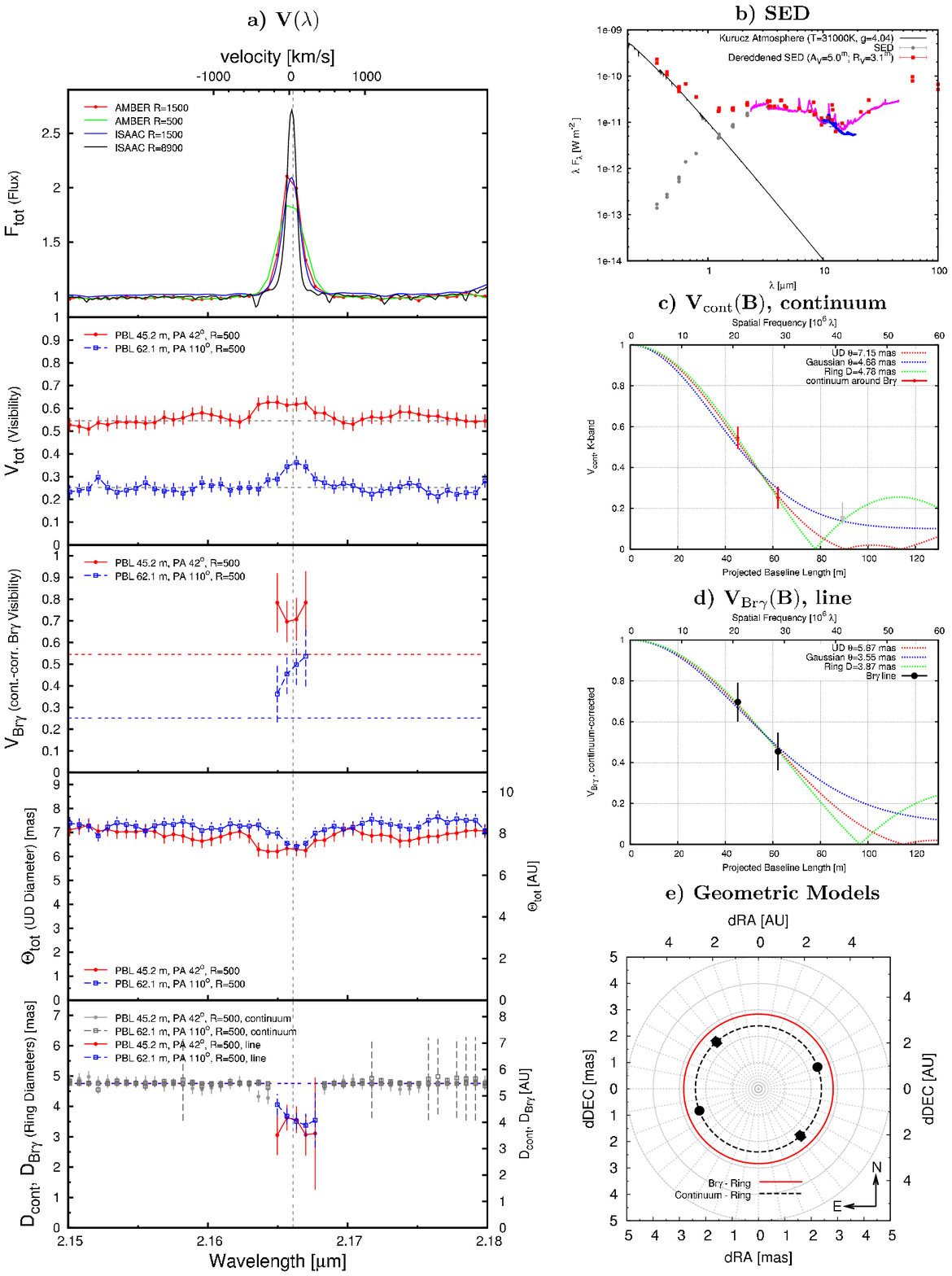}\\
  \caption{Spectroscopic, spectro-interferometric, and photometric data for
    V921\,Sco (similar to Fig.~\ref{fig:hd104237}). }
  \label{fig:v921sco}
\end{figure*}

The wavelength-dependent visibilities extracted from our AMBER data are shown
in the second panel from the top in Figs.~\ref{fig:hd104237}{\it(a)} to
\ref{fig:v921sco}{\it(a)}.  Four stars in our sample (HD\,163296, HD\,104237,
HD\,98922, V921\,Sco) show a visibility increase within the {\BrG} line.  For
MWC\,297, the visibility drops within the line, as already reported by
\citet{mal07}.  

Since the spectral channels which include the {\BrG} line additionally contain
flux contributions from the photosphere and circumstellar material, the
interpretation of the measured wavelength-dependent visibilities requires
quantitative modeling, as presented in Sect.~\ref{sec:modelingcont}.

\subsection{Closure phases}
\label{sec:closurephases}

\begin{table*}[t]
  \caption{Closure Phases extracted from the data. }
  \label{tab:closurephases}
  \centering

\begin{tabular}{l|cc|cc|c|c}
  \hline\hline
  & \multicolumn{4}{c|}{Projected Baselines} & \multicolumn{2}{c}{Closure Phase}\\
  & $u_{12}$     & $v_{12}$         & $u_{23}$ & $v_{23}$       & continuum  & $\mathrm{Br}\gamma$ line \\
  & [m]         & [m]              & [m]     & [m]            & [deg]            & [deg]\\
  \noalign{\smallskip}
  \hline 
  \noalign{\smallskip}
HD\,104237 &
  33.1 & 11.6 &     52.9 & -30.0 &      $9.8 \pm 6.7$ & --- \\
HD\,163296 &
  14.5 & 41.4 &     51.6 & -8.0 &      $1.5 \pm 4.3$ & --- \\
HD\,98922 &
  30.9 & 29.7 &     57.6 & -22.6 &      $-9 \pm 16$ & --- \\
V921\,Sco &
  30.2 & 33.7 &     58.3 & -21.6 &      $20 \pm 24$ & $3.1 \pm 2.8$ \\
  \noalign{\smallskip}
  \hline
\end{tabular}
\end{table*}

Closure phase measurements can provide unique information about deviations 
from centro-symmetry in the source brightness distribution.
For YSO disk geometries, such asymmetries are expected in particular for systems 
seen under intermediate inclination. 
The strongest CP signals are predicted by disk models with vertical puffed-up inner rim
\citep[e.g.~][]{dul01}, while models with curved inner rims predict a smoother,
more symmetric brightness distribution corresponding to smaller CPs \citep[e.g.~][]{ise05}.

Given the importance of this observable, we present here the CPs extracted
from our data, although the error bars on most measurements are rather large, 
typically due to the low fringe contrast on the longest baseline in the
telescope triplet.
In Tab.~\ref{tab:closurephases} we list the CPs measured from our datasets
and give the corresponding $(u, v)$ baseline vector for two baselines in 
the employed telescope triplet (the third $(u, v)$-vector is given by the closure relation).
We find that all continuum CPs are consistent with a zero CP on the $1\sigma$ (HD\,163296, HD\,98922, V921\,Sco) 
or $2\sigma$ level (HD\,104237).
This adds support to the conclusions drawn by \citet{mon06}, who used the IOTA-3T
interferometer (providing baseline lengths up to 38~m) and measured for 12 out of 13
sources CPs below $\sim 5${\degr} (excluding one binary source).
Since the CP signal is expected to increase rapidly with baseline lengths, 
our new observations (with baselines up to 89~m) provide new constraints on this issue
and favour again a centro-symmetric brightness distribution.
In Sect.~\ref{sec:interpretation}, we discuss some of our CP measurements qualitatively, 
but leave it to future studies to investigate whether our results are also in 
quantitative agreement with the current generation of rim models.

For V921\,Sco, the particularly strong {\BrG} line flux allowed us also to measure
an accurate CP measurement within the spectrally resolved emission line, 
yielding a CP signal of $3.1 \pm 2.8${\degr}, consistent again with a 
centro-symmetric brightness distribution.

\section{Modeling: Continuum-emission}
\label{sec:modelingcont}

\begin{table*}[t]
  \caption{Best-fit parameters for our geometric model fits.}
  \label{tab:geomodelparameters}
  \centering

\begin{tabular}{l|cc|cc|cc|cc|cc}
  \hline\hline
                   & \multicolumn{6}{c|}{Continuum emission}    & \multicolumn{2}{c|}{$\boldmath{\mathrm{Br}\gamma}$ emission} & \multicolumn{2}{c}{$\boldmath{\mathrm{Br}\gamma}$ emission} \\
                   & &&&&&& \multicolumn{2}{c|}{line center} & \multicolumn{2}{c}{all spectral channel} \\
                   & \multicolumn{2}{c|}{Gaussian} & \multicolumn{2}{c|}{Uniform Disk}       & \multicolumn{2}{c|}{Ring} & \multicolumn{2}{c|}{Ring} & \multicolumn{2}{c}{Ring} \\
                   & FWHM$_{\mathrm{cont}}$ & $\chi^2_{\mathrm{red}}$ & $\Theta_{\mathrm{cont}}$ & $\chi^2_{\mathrm{red}}$ & $D_{\mathrm{cont}}$ & $\chi^2_{\mathrm{red}}$ & $D_{\mathrm{Br}\gamma}$ & $\chi^2_{\mathrm{red}}$ & $D_{\mathrm{Br}\gamma}$ & $\chi^2_{\mathrm{red}}$ \\
  Target Star      & [AU]                &              & [AU]              &              & [AU]             &              & [AU]         &               & [AU]         & \\
  \noalign{\smallskip}
  \hline
  \noalign{\smallskip}
HD\,104237 &
$0.57 \pm 0.10$ &  1.0 &
$0.87 \pm 0.08$ &  0.5 &
$0.58 \pm 0.04$ &  0.3 &
$0.31 \pm 0.13$ &  0.6 &
$0.35 \pm 0.21$ &  0.5 \\
HD\,163296 &
$0.39 \pm 0.05$ &  0.9 &
$0.61 \pm 0.08$ &  1.5 &
$0.41 \pm 0.06$ &  1.9 &
$0.35 \pm 0.15$ &  0.1 &
$0.25 \pm 0.19$ &  0.1 \\
HD\,98922 &
$2.45 \pm 0.26$ &  0.5 &
$3.70 \pm 0.27$ &  0.2 &
$2.47 \pm 0.16$ &  0.2 &
$< 0.5$ & --- &
$< 0.5$ & --- \\
MWC\,297 &
$1.27 \pm 0.10$ &  --- &
$1.99 \pm 0.13$ &  --- &
$1.35 \pm 0.08$ &  --- &
$1.89 \pm 0.52$ &  0.1 &
$2.52 \pm 0.44$ &  0.4 \\
V921\,Sco &
$5.38 \pm 0.52$ &  0.7 &
$8.22 \pm 0.55$ &  0.1 &
$5.49 \pm 0.32$ &  0.1 &
$4.45 \pm 0.55$ &  0.1 &
$4.35 \pm 0.64$ &  0.2 \\
  \noalign{\smallskip}
  \hline
\end{tabular}

\begin{flushleft}
  \hspace{5mm}Notes~--~$\chi^2_{\mathrm{red}}$ is defined as $\sum \left[(V_{\mathrm{meas}}-V_{\mathrm{model}})/\sigma_{V_{\mathrm{meas}}}\right]^{2}/(N-1)$, where
    $N$ is the number of measurements, $V_{\mathrm{model}}$ is the model visibility, and $V_{\mathrm{meas}}$
    and $\sigma_{V_{\mathrm{meas}}}$ are the measured visibility and total error, respectively.\\
\end{flushleft}
\end{table*}

Over the last few years, ring models with uniform brightness have emerged
as the prototypical geometry for the interpretation of YSO interferometric data.
In most cases, this prefered use of ring geometries instead of other simple geometries
(e.g.\ uniform disk, Gaussian, or disk geometries with constant temperature power-law) 
is not explicitly required to reproduce the details of the sampled visibility function
(which typically covers only the first lobe and is
thus rather insensitive to the inner gap in the brightness distribution), 
but mainly based on indirect evidence or theoretical arguments.
For example, some of the first infrared interferometric measurements on HAeBe stars
\citep[e.g.\ ][]{mil01} already showed that classical geometrically thin accretion disk 
geometries extending down to several stellar radii might be consistent with 
interferometric measurements for early-type Herbig~Be stars, but result in too 
compact structures for Herbig~Ae and late-type Herbig~Be stars \citep{eis04,vin07}.
Ring geometries, on the other hand, yield sizes which are consistent with the
expected dust sublimation radii and follow the predicted stellar luminosity 
scaling law \citep{mon02}.  
In the meantime, this finding was also interpreted in theoretical
work \citep[e.g.\ ][]{nat01,dul01,ise05}, attributing the near-infrared
continuum emission mainly to hot dust located at the dust sublimation radius.
Based on these arguments, we also prefer to use ring-like geometries 
for the interpretation of our visibility data
(assuming a fractional ring width of 20\%, \citealt{mon05}), but also give uniform disk (UD) and Gaussian
FWHM diameters to allow comparison with other work.
In order to estimate the contribution of the stellar photosphere to the
total $K$-band flux, we estimate the flux ratio using the SED and the Kurucz
atmosphere models shown in Figs.~\ref{fig:hd104237}{\it(b)} to \ref{fig:v921sco}{\it(b)}.
For more details about the model fitting procedure, we refer to one of our earlier
studies \citep{kra08}.

Besides model fits to each individual spectral channel (see fourth and fifth
panel of Figs.~\ref{fig:hd104237}{\it(a)} to \ref{fig:v921sco}{\it(a)}), we also fitted these
models to continuum-visibilities, which were averaged over several
spectral channels around 2.15 and 2.18~$\mu$m and plot the corresponding
measurements and model curves as a function of baseline length
(Figs.~\ref{fig:hd104237}{\it(c)} to \ref{fig:v921sco}{\it(c)}).  
In these plots, we also show the upper limits, which we can put on the
continuum visibility using the baselines which had to be rejected due to low SNR
(see Sect.~\ref{sec:obsamber}).
The fitted diameters for Gaussian, UD, and ring profiles are listed in
Tab.~\ref{tab:geomodelparameters}.
To convert the measured angular size to physical scales, we assume the distances listed in Tab.~\ref{tab:targetstars}.
Please note that the errors given do not take distance uncertainties into account.
By fitting the visibilities measured on different baseline orientations simultaneously, 
we assume that the intensity profile does not depend on position angle (e.g.\ face-on disk).
Since the visibilities derived for HD\,163296 provide sufficient
position angle coverage to investigate for a possible elongation of the 
continuum-emitting region (e.g., due to disk inclination), we also fitted inclined 
ring geometries to this data set and derived an inclination of 
$68\pm10$\degr\ with a PA of $144\pm9$\degr\ (corresponding to an inclined ring diameter 
of $1.02 \times 0.37$~AU).

\section{Modeling: $\boldmath{\mathrm{Br}\gamma}$ line emission}
\label{sec:modelingbrg}

From each spectral channel of our AMBER interferograms, we can derive a value
for the visibility amplitude, providing spatial information about the
brightness distribution contributing to this spectral channel.
Within the {\BrG} line, the flux within a spectral channel is composed of the
line emission plus the underlying continuum contribution.  To model such
visibility data, either the composite (line \& continuum) object can be
modeled, or the measured continuum+line visibility has to be corrected for the
continuum contribution, yielding the visibility of the pure line-emitting
region.

We correct the measured visibilities using the relation
\begin{equation}
  V_{\mathrm{Br}\gamma}=\frac{F_{\mathrm{tot}} V_{\mathrm{tot}} - F_{\mathrm{cont}} V_{\mathrm{cont}}}{F_{\mathrm{Br}\gamma}}, \label{eqn:VBrGcontcorr}
\end{equation}
where $F_{\mathrm{tot}}$ and $V_{\mathrm{tot}}$ are the flux and the
visibility measured within an arbitrary spectral channel. 
To compute the {\BrG}-line flux 
$F_{\mathrm{Br}\gamma}:=F_{\mathrm{tot}}-F_{\mathrm{cont}}$, we
determine the continuum flux $F_{\mathrm{cont}}$ from Kurucz atmosphere models,
taking the underlying photospheric {\BrG} absorption component into account.
Then we interpolate the 
continuum visibility $V_{\mathrm{cont}}$ over the {\BrG}~line and derive the
continuum-corrected visibility $V_{\mathrm{Br}\gamma}$ of the {\BrG}-emitting
region. As discussed in \citet[][ Appendix~C]{wei07}, this relation assumes a
negligible (zero) differential phase (corresponding to coinciding
photocenters between the continuum- and line-emitting regions).
We compute the continuum-correction for each spectral channel of the AMBER data where
$F_{\mathrm{Br}\gamma}$ is sufficiently large to apply Eq.~\ref{eqn:VBrGcontcorr}
reliably and show the resulting $V_{\mathrm{Br}\gamma}$ values in the third panel from
the top of Figs.~\ref{fig:hd104237}{\it(a)} to \ref{fig:v921sco}{\it(a)}.

The interpretation of the derived line visibilities is difficult for several reasons.
For instance, it is likely that the geometry of the line-emitting
region is more complex than the continuum-emitting region, possibly
extending above the equatorial disk plane, perhaps introducing strong
inclination effects.  Furthermore, as indicated by the short-period line
variability detected towards YSOs of all masses, the kinematics and possibly also
morphology of the line-emitting gas might change significantly even on
short time scales of days or weeks, involving, for example, the ejection of
fast-moving blobs at the base of an outflow.  Finally, it is likely that
several processes (including accretion and outflow processes) contribute to
the {\BrG} emission.
Given this complexity and the large error bars, we refrain from applying complex
line radiative transfer models for the interpretation of the measured line
visibilities in this first paper and instead limit ourselves to fitting simple ring-geometries,
assuming one dominant source for the {\BrG} emission.
Since the derived line visibility values are rather
high for most objects, the characteristic size of the {\BrG}-emitting region does not
critically depend on the intensity profile of the underlying brightness
distribution and therefore affects our conclusions only marginally.
Since the {\BrG} line was spectrally resolved by AMBER in all observations, we
performed two fits: one including all spectral channels where $V_{\mathrm{Br}\gamma}$ could
be determined and one including only the spectral channel at line
center.
The circular-symmetric ring-model fits to the continuum-corrected {\BrG} visibilities are
shown in Figs.~\ref{fig:hd104237}{\it(d)} to \ref{fig:v921sco}{\it(d)}, and the determined
best-fit ring diameters are listed in Tab.~\ref{tab:geomodelparameters}.

\section{Discussion of individual objects}
\label{sec:interpretation}

\subsection{HD\,104237: A circumbinary disk}
\label{sec:hd104237}

HD\,104237 (alias DX~Cha, CD-77$^{\circ}$528, Hen~3-741) has an approximate age
of $\sim2$~Myrs \citep{boe04}.
Imaging observations by \citet{gra04} suggest that HD\,104237 is associated
with a microjet (HH\,669) and that the disk is seen nearly face-on
($i=18^{+14}_{-11}\,^{\circ}$).  
Spectroscopic monitoring suggests that HD\,104237 consists of a binary system
on a short-period (19.859~d), highly eccentric ($e \approx 0.66$) orbit with
component masses of $M_{A}=2.2\pm 0.1$~M$_{\sun}$ and $M_{B}=1.7\pm
0.1$~M$_{\sun}$ \citep{boe04}.
Using these orbital elements, we estimate the major axis of the binary system
to be $\sim 1.91$~mas (0.23~AU), which places both stellar components inside of the 
ring-diameter measured by our AMBER observations at continuum wavelengths
($\sim 5.0$~mas~=~0.58~AU).
The measured continuum ring radius of 0.29~AU is very close to the expected dust
sublimation radius (0.32~AU, $L=30~L_{\sun}$, $T_{\mathrm{subl}}=1500$~K, see 
Sect.~\ref{sec:discusscont}), suggesting that the continuum emission traces 
mainly dust located at the inner rim of the circumbinary dust disk.

In the {\BrG} line, we measured no significant deviation of the line visibility with
respect to the continuum visibility on 2005-02-26 at a short baseline length
($B_{p} \sim 35$~m), while on 2007-01-09, the new measurements at $B_{p} \approx 58$~m show a
significant increase in visibility (see Fig.~\ref{fig:hd104237}). In both
epochs, the continuum-corrected {\BrG}-line visibility indicates that the
emitting region has a similar or smaller extension than the continuum-emitting
region. 
In all available measurements, the {\BrG}-emitting region is significantly
more extended than the co-rotation radius, which leads us to reject the
magnetospheric accretion scenario and favor the disk wind scenario 
(consistent with the conclusion from \citealt{tat07a}).
Due to the low mass accretion rate of this object \citep[$\log \dot{M}_{\mathrm{acc}}
= -7.45$,][]{gar06}, we deem major line contributions from a gaseous
inner disk unlikely \citep{muz04}.

For each epoch, the data cover only single baseline lengths, which makes
it difficult to estimate the influence of the binary companion both on the
continuum and the {\BrG} visibilities.
Given our marginal detection of a non-zero continuum closure phase signal for HD\,104237
(see Tab.~\ref{tab:closurephases}) and the estimated binary parameters,
it seems likely that we also detect small contributions from the inner
resolved binary system.
In particular, the binary nature might result in a significant asymmetry in
the brightness distribution (in continuum and/or line emission) which could
bias both the correction for the stellar contribution (see
Sect.~\ref{sec:SED}) as well as the continuum-correction 
required to compute $V_{\mathrm{Br}\gamma}$ (see Sect.~\ref{sec:modelingbrg} and
Eq.~\ref{eqn:VBrGcontcorr}).
Therefore, future investigations on this object will require a significantly
larger amount of observational data to reliably discern the influence of these
effects.

\subsection{HD\,163296}
\label{sec:mwc275}

HD\,163296 (alias MWC\,275) is a particularly well-studied Herbig~Ae star.
CO millimeter line observations have revealed the outer parts of a Keplerian
rotating disk \citep{man97,ise07}.
The determined disk inclination angle ($i=46 \pm 4\,^{\circ}$) as well as the
position angle (PA=$128\pm4\,^{\circ}$) are in reasonable agreement with the orientation of
a disk-like structure seen in scattered light from coronographic imaging
\citep[$i=60\pm5^\circ$, PA=$140 \pm 5^\circ$,][]{gra00}.
Imaging observations have revealed an asymmetric, bipolar jet emerging
roughly perpendicular to the disk plane and terminating in the Herbig-Haro
objects HH\,409-A to C, which were found along PA=$42.5\pm 3.5^\circ$.
The inclination angle inferred from the jet observations is
$i=51^{+11}_{-9}\,^{\circ}$ \citep{was06}.

The infrared continuum environment around HD\,163296 was already resolved with 
the VLTI/MIDI ($N$-band; $V(99\mathrm{m})\approx 0.2$; \citealt{lei04}), IOTA
($H$-band; $V(38\mathrm{m})\approx 0.85$; \citealt{mil01}), and the Keck interferometer
($K$-band; $V(76\mathrm{m})\approx 0.47$; \citealt{mon05}).
Recent measurements obtained with the CHARA interferometer \citep{tan08}
indicate that a significant fraction of the NIR continuum emission emerges 
from inside the dust sublimation radius, maybe due to gas contributions.

High-resolution spectroscopy ($R=25\,000$) of the {\BrG} line of
HD\,163296 was presented by \citet{bri07}, showing a single-peaked line profile 
consistent with our ISAAC spectra.
For the origin of the Balmer lines, \citet{pog94} proposed a stellar
wind/outer shell model.

In the $V(B)$-plot for the continuum wavelength regime
(Fig.~\ref{fig:mwc275}{\it(c)}), our AMBER measurements towards different baseline position angles
(covering a PA range of 80\degr)
cannot be well represented with the visibility profiles corresponding to 
ring-, UD-, or Gaussian-geometries, which might indicate either a very unusual radial
intensity profile or, more likely, an elongation of the continuum-emitting
region.
The latter hypothesis can be investigated with the inclined ring model fits shown in
Fig.~\ref{fig:mwc275}{\it(e)}, indicating in fact that the near-infrared emitting region is strongly
elongated.  The major axis of the continuum-emitting region (PA~$144 \pm 9$\degr) seems to be oriented
perpendicular to the outflow direction indicated by the optical microjet \citep[PA~$42.5\pm 3.5^\circ$,][]{was06}.
Due to the limited position angle coverage of our interferometric observations, the disk inclination angle 
($i=68$\degr) is only poorly constrained by our observations, but seems to be in general in good agreement
with the high inclination inferred from earlier observations.
In spite of the strong inclination, the derived continuum closure phase for HD\,163296 is still surprisingly 
close to zero (see Tab.~\ref{tab:closurephases}), favouring disk models without notable inclination-induced skews.

As indicated by the quite flat wavelength-dependent visibility curve
(Fig.~\ref{fig:mwc275}{\it(a)}, second panel), the {\BrG} emission originates
in a region of similar spatial extent as the continuum-emitting region.
On the longer baselines, we find weak evidence ($\sim 2\sigma$) for an
increase of visibility within the {\BrG} line, which indicates that most of
the line emission emerges from within the dust-sublimation radius.  
Plotting the continuum-corrected {\BrG}-line visibility as a function of
baseline length shows that the {\BrG}-emitting region is likely spatially
resolved (with $R_{\mathrm{Br}\gamma}/R_{\mathrm{cont}} \approx 0.7$), which
leads us to favor a stellar wind or a disk wind as the emitting mechanism.   
Considering the low mass accretion rate of HD\,163296 
\citep[$\log \dot{M}_{\mathrm{acc}} = -7.12$,][]{gar06}, 
it seems unlikely that
the {\BrG} line emission emerges from a gaseous inner disk \citep{muz04}.

\subsection{HD\,98922}
\label{sec:hd98922}

For the B9Ve-type star HD\,98922, a moderate infrared excess 
\citep{mal98b} and a very high mass accretion rate was
determined (\citealt{gar06}; $\log \dot{M}_{\mathrm{\mathrm{acc}}} =
-5.76~M_{\sun}$\,yr$^{-1}$, which is the highest mass accretion rate in the
sample of 36 Herbig~Ae stars investigated by these authors).
Outflow signatures were detected in the {\Ha} as well as in the {\SiII} line
($v \approx 300$~km\,s$^{-1}$, \citealt{gra96}). 
\citet{ack05} tried to model the profile of the {\ForbOI} emission line
with a Keplerian rotating disk model and concluded that the emission likely 
does not originate in the disk surface
as in several other sources, but possibly in a rotating gaseous disk inside
the dust-sublimation radius. Van Boekel et al.\ (\citeyear{van03}) studied
the profile of the 10\,$\mu$m~silicate emission feature detected towards
HD~98922 and found spectral features associated with olivines and pyroxenes,
indicating the presence of large, rather evolved grains.

The inclination under which HD\,98922 is seen is still not well known.
\citet{blo06} modeled the UV spectrum and derived an intermediate inclination angle
($i \approx 45$).  
Since the ring-diameters which we derive from our AMBER measurements are nearly
identical towards very different position angles 
(Fig.~\ref{fig:hd98922}{\it(a)}, lowest panel), we find no indications for a
large disk inclination, but cannot exclude it either.
Despite the fact that HD\,98922 has the lowest observed {\BrG} line equivalent width 
of the stars in our sample ($F_{\mathrm{Br}\gamma}/F_{\mathrm{cont}} \approx
1.2$), we measure a strong increase of visibility within the {\BrG} line with
respect to the continuum visibility (see Fig.~\ref{fig:hd98922}).  Remarkably,
applying the continuum-correction to the measured {\BrG}-line visibility
reveals that the {\BrG} line-emitting region seems to be completely unresolved
(i.e.\ $V_{\mathrm{Br}\gamma} \approx 1$) at baseline lengths up to $\sim 60$~m,
while the continuum-emitting region is resolved with a 
ring-diameter of $4.6 \pm 0.1$~mas ($2.48 \pm 0.05$~AU).
Therefore, the {\BrG}-emitting region is at least a factor of 5 times more
compact than the dust-sublimation radius, which is consistent with the stellar
wind, X-wind, and magnetospheric accretion scenario.
Given the spectroscopic indications for an exceptionally high mass accretion
rate, we favor the magnetospheric accretion scenario.

\subsection{MWC\,297}
\label{sec:mwc297}

MWC\,297 (alias NZ~Ser) is a B1.5Ve-type Herbig~Be star with a remarkably high rotation
velocity of $350 \pm 50$~km\,s$^{-1}$ (as derived from photospheric lines),
suggesting that the system might be seen nearly edge-on \citep{dre97}.
On the other hand, using spectro-polarimetry in the {\Ha} line, \citet{oud99}
could not find any evidence for asymmetry in the {\BrG} line-emitting region,
as would be expected for a nearly edge-on system.
\citet{mal07} applied an analytic accretion disk model and could reproduce the
SED and AMBER $K$-band and IOTA $H$-band visibilities with an inclination
angle of $i \sim 20\deg$ (i.e.\ closer to face-on than edge-on).
On the other hand, \citet{ack08} investigated the near- and mid-infrared 
geometry of the circumstellar environment around MWC\,297 and found
best agreement using a two-component Gaussian geometric model.

The hydrogen line emission observed towards MWC\,297 is exceptionally strong
and has been interpreted as stellar wind \citep{nis95,mal07} or 
accretion-driven mass-loss from the disk \citep{cor98}.

The UKIRT spectra of \citet{dre97} show a single-peaked profile for the
{\Ha}, {\Hb}, and {\BrA} line.  \cite{gar06} presented an ISAAC spectrum of
MWC\,297, showing a double-peaked {\BrG}-line profile.  Based on our 
re-reduction of the ISAAC spectroscopic data, we cannot confirm the
double-peaked profile of the {\BrG} line, but classify the line profile as
single-peaked with the currently available resolution ($R\sim9000$). NIR
spectroscopy with higher spectral resolution will be required to determine the
real underlying line profile.

As already discussed by various other authors \citep{mon05,mal07}, the
continuum visibilities measured on MWC\,297 indicate a very compact
continuum-emitting region, corresponding to a ring-radius approximately
5-times more compact than the dust sublimation radius expected for an
irradiated dust disk.  The interpretation of this effect (which was also observed
towards several other Herbig~Be stars) is still debated, but might
include either gas absorption \citep[allowing dust to exist closer to the star; e.g.\ ][]{mon02,mal07} 
or the emission of an optically thick gaseous inner accretion disk
\citep[e.g.\ ][]{mon05,kra08}. 

For MWC\,297, the visibility drops significantly within the {\BrG} line (see
Fig.~\ref{fig:mwc297}), indicating that the line-emitting region is more
extended than the continuum-emitting region.  
\citealt{mal07} modeled this observational result assuming an optically thick
gas disk with an inner radius of 0.5~AU and an outflowing stellar
wind.
Besides this interpretation, a disk wind scenario would likely also be consistent
with our data.
In this context, it is important to note that the pronounced drop of
visibility within the {\BrG} line is also supported by the particular
compactness of the continuum-emitting region of MWC\,297.
As discussed above, the continuum-emission around MWC\,297 might not be
dominated by pure dust emission, but instead by the emission of gas+dust
located close to the star.
Therefore, the observation that MWC\,297 shows a drop of visibility within the 
{\BrG} line, while V921\,Sco (which has a similar spectral type as MWC\,297)
shows an increase of visibility, might not indicate substantially different
line formation mechanisms, but may simply reflect different physical
conditions in the continuum-emitting dust and gas disk (e.g.\ optically thick
vs.\ optically thin gas disk).

In case of a nearly edge-on system inclination, the influence of geometric
effects (extension of the {\BrG}-emission region above the mid-plane,
appearance of bipolar lobes, etc.) would also be of importance for the
interpretation and modeling of the spectro-interferometric data on MWC\,297.

\subsection{V921\,Sco}
\label{sec:v921sco}

In the catalog of HAeBe member and candidate stars by \citet{deW90},
V921\,Sco (alias MWC\,865, CD-42$^{\circ}$11721, Hen~3-1300) is classified as
a B0[e]p-type star, although there is some dispute both about the stellar
parameters and the evolutionary stage of this object (see discussion in
\citealt{bor07}).  In the vicinity of V921\,Sco, a cluster of YSO
candidate sources were detected \citep{hab03, wan07}, and the 
star is also associated with an extended nebulosity, which
was observed at optical \citep{van75,hut90}, near-infrared \citep{wan07},
mid-infrared \citep{nat93}, as well as sub-mm wavelengths \citep{hen98}. 
Comparing the spectral slope of the ISO spectrum with that of the {\it
  Spitzer}/IRS spectrum taken with a smaller aperture (see
Fig.~\ref{fig:v921sco}b) suggests that the SED at $\lambda \gtrsim 15~\mu$m is
dominated by emission from this envelope.
The SED model fits by \citet{bor07} suggest that the system
is seen under low to intermediate inclination ($i \lesssim 70$\degr).

The line profiles of the strong hydrogen recombination lines were
modeled as emission from a spherically symmetric gas envelope
\citep{ben98}. 
\citet{ack05} interpreted the profile of the optical
{\ForbOI} line in the context of a wind originating from the surface layer
of a passive disk.

Using the AMBER visibilities measured at continuum wavelengths, we compare the
ring diameter measured towards various PAs (covering a position angle
range of $\sim 68\deg$) and find no indications for an elongation of the
continuum-emitting region.
This is consistent with the measured small {\BrG} line closure phase (see Tab.~\ref{tab:closurephases}), 
indicating that the brightness distribution of the combined line- and continuum-emitting region
is nearly centro-symmetric, as in the case when the line-emitting region is 
seen nearly face-on.
Furthermore, V921\,Sco exhibits a continuum-emitting region which is more compact than 
expected for an irradiated dust disk, which might suggest that the NIR continuum 
emission is dominated by emission from an optically thick gaseous inner accretion disk 
(similar to MWC\,297).

Within the {\BrG} line, we measure a slight increase in visibility
(see Fig.~\ref{fig:v921sco}).  Furthermore, the continuum-corrected
line-visibilities $V_{\mathrm{Br}\gamma}$ (ranging between 0.7 and 0.5) show
that the {\BrG}-emitting region is also spatially resolved and only slightly
more compact than the continuum-emitting region
($R_{\mathrm{Br}\gamma}/R_{\mathrm{cont}} \approx 0.7$).
Therefore, the {\BrG} region is too extended to be consistent with
magnetospheric accretion or an X-wind as dominant {\BrG}-emitting mechanism,
which makes a strong stellar wind, a disk wind or a gaseous inner disk the
most likely scenario.

\section{Discussion of general trends}
\label{sec:discussion}

\begin{figure}[htbp]
  \centering
  \includegraphics[angle=270,width=8cm]{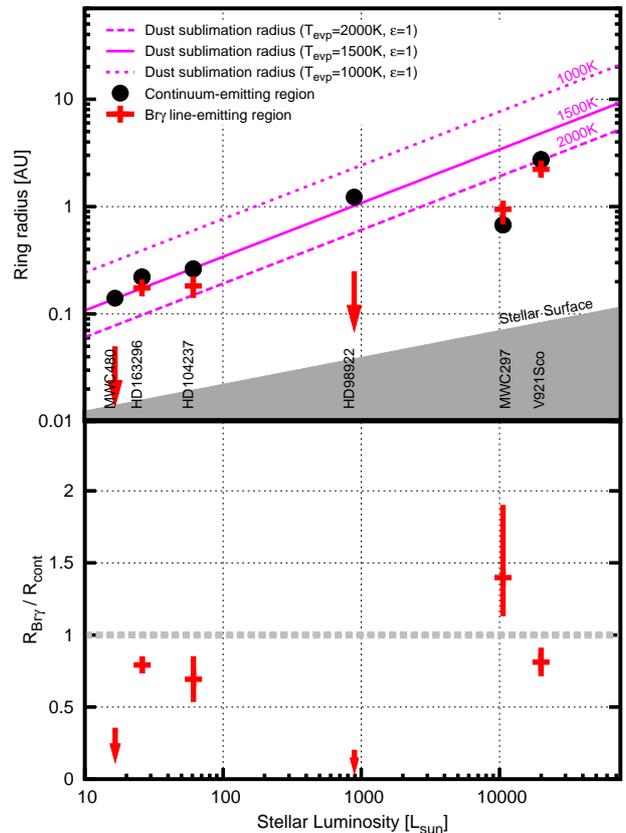}\\[2mm]
  \caption{{\it Top:} The fitted ring radii for the continuum ($R_{\mathrm{cont}}$, black points) and
    {\BrG} line ($R_{\mathrm{Br}\gamma}$, red points) plotted as a function of stellar luminosity.
    For the {\BrG} line, we show the ring radius, determined by fitting only
    the spectral channel with the line center.
    For comparison, we plot the spatial extension of the stellar surface $R_{\star}$ (grey area) 
    and the dust sublimation radius corresponding to dust sublimation 
    temperatures of 2000, 1500, and 1000~K (computed using Eq.~\ref{eqn:sublradiusIN05} 
    and assuming $\epsilon=1$).
    {\it Bottom:} Providing a natural measure for the temperature distribution
    within the circumstellar disk, we normalized $R_{\mathrm{Br}\gamma}$ by the size
    of the continuum-emitting region $R_{\mathrm{cont}}$.
    In particular, this normalization seems important considering the large
    range of stellar luminosities covered by our sample.
 }
  \label{fig:RRatiovsLstar}
\end{figure}

\begin{figure}[htbp]
  \centering
  \includegraphics[angle=270,width=8cm]{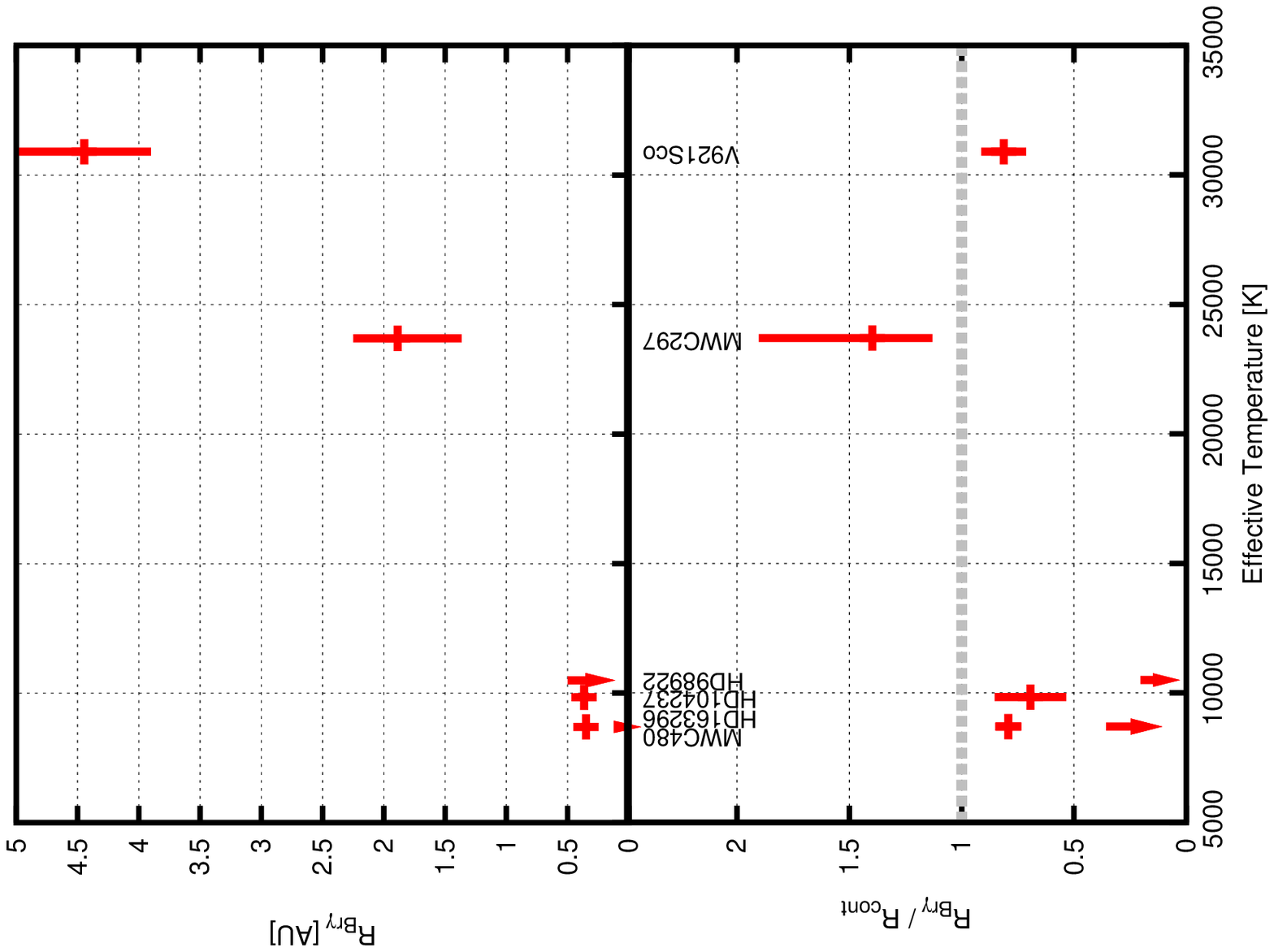}\\[5mm]
  \caption{$R_{\mathrm{Br}\gamma}$ and $R_{\mathrm{Br}\gamma}/R_{\mathrm{cont}}$ plotted as a function of effective stellar temperature.}
  \label{fig:RRatiovsTeff}
\end{figure}

\begin{figure}[htbp]
  \centering
  \includegraphics[angle=270,width=8cm]{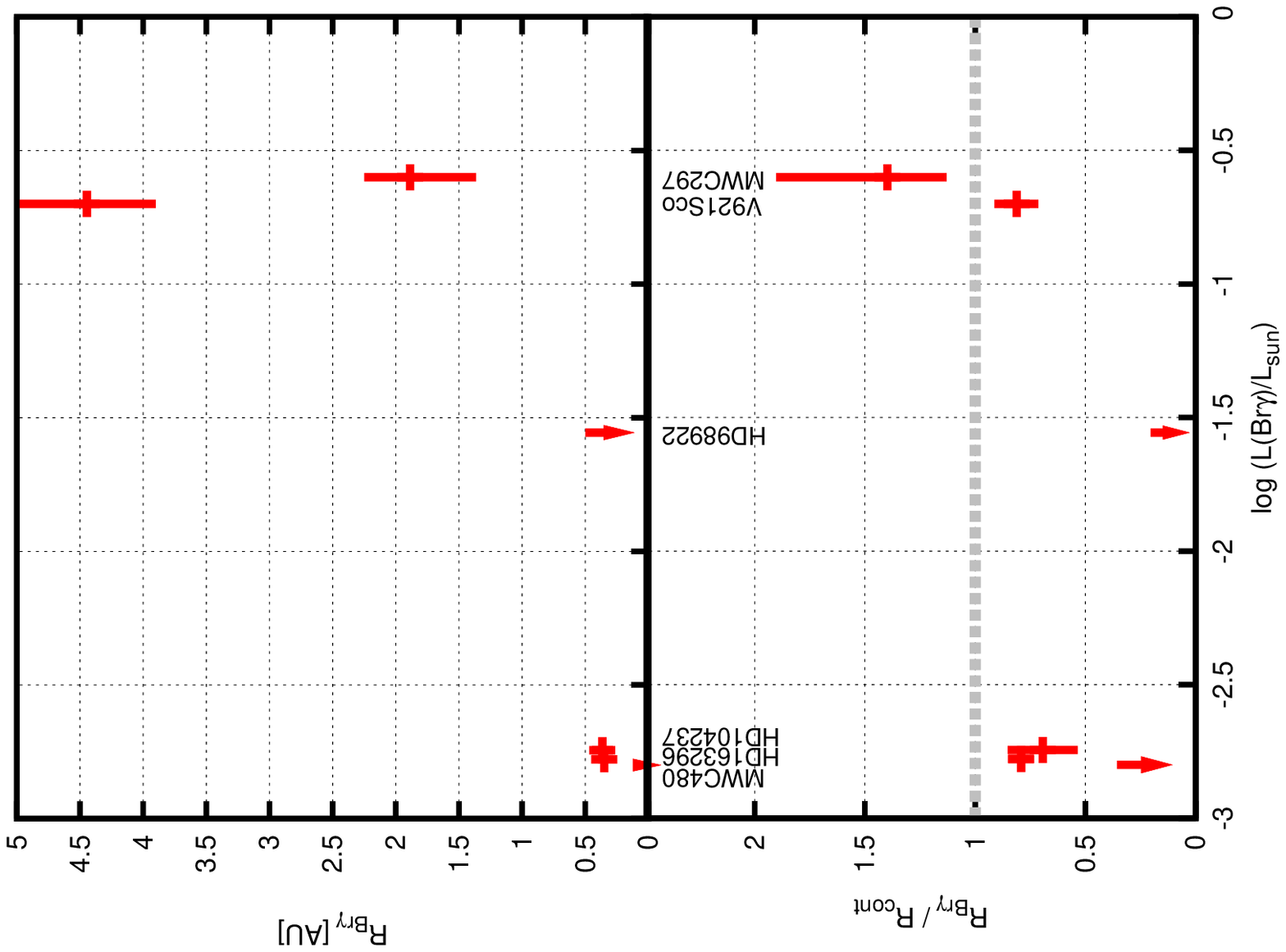}\\[5mm]
  \caption{$R_{\mathrm{Br}\gamma}$ and $R_{\mathrm{Br}\gamma}/R_{\mathrm{cont}}$ plotted as a
    function of the {\BrG} line luminosity.}
  \label{fig:RRatiovsLBrG}
\end{figure}

\begin{figure}[htbp]
  \centering
  \includegraphics[angle=270,width=8cm]{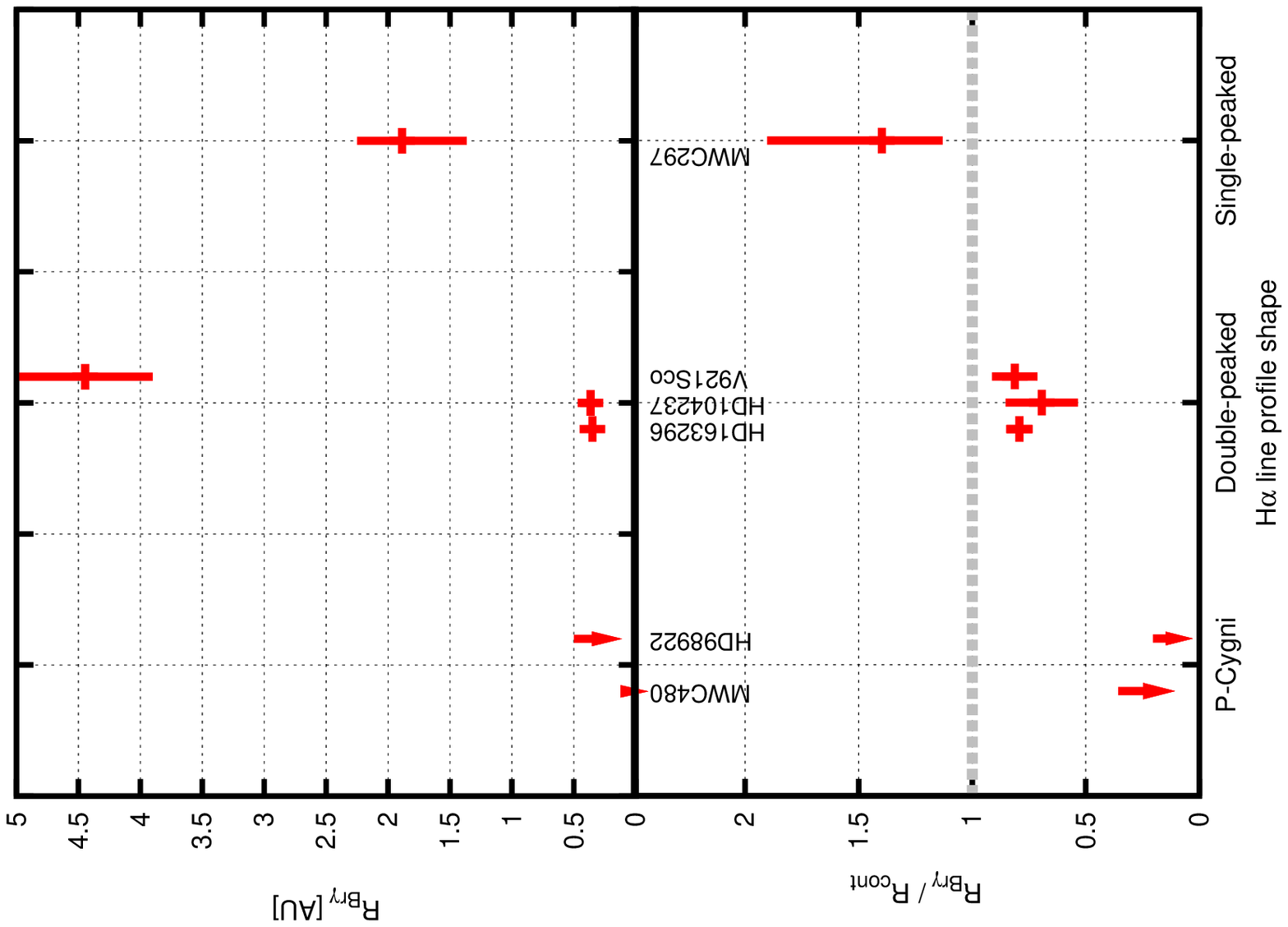}\\[5mm]
  \caption{$R_{\mathrm{Br}\gamma}$ and $R_{\mathrm{Br}\gamma}/R_{\mathrm{cont}}$ versus the {\Ha}~line profile
    (see Tab.~\ref{tab:targetstars}).}
  \label{fig:RRatiovsPHa}
\end{figure}

Since our sample covers a wide range of stellar parameters, we investigated
whether the measured size of the continuum and {\BrG}-emitting region
correlates with the stellar parameters or the spectroscopic properties.
Revealing such relationships could provide more insight 
into the involved physical mechanisms and is also essential to confirm the empirically found
correlations between the {\BrG} luminosity and other estimators for the mass
accretion rate \citep{cal04}.
In order to further expand our sample, we include in this section not only the
five objects which we have investigated with AMBER, but also the
MWC\,480 Keck-Interferometer measurement published by \citet{eis07b}.  For
MWC\,480, we assume $L=17~L_{\sun}$, $d=140$~pc, $T_{\mathrm{eff}}=8700$~K, a
P-Cygni {\Ha}-line profile \citep{ack05}, and a {\BrG} line luminosity 
$L($\BrG$)/L_{\sun}=-2.8$ (as determined by L.\ Testi and A.\ Natta from 
unpublished TNG spectra; private communication).

\subsection{Continuum emission}
\label{sec:discusscont}

Earlier interferometric investigations of the continuum-emitting region
of HAeBe stars have established a relation between the size of the
continuum-emitting region and the stellar luminosity, suggesting that the
continuum emission mainly traces hot dust located at the dust sublimation
radius.
In order to check for similar trends, we plot the determined $K$-band
continuum ring radii as a function of the stellar
luminosity $L_{\star}$ in Fig.~\ref{fig:RRatiovsLstar}.  
For comparison, we computed the predicted inner disk
radii corresponding to dust sublimation temperatures $T_{\mathrm{subl}}$ of
2000~K, 1500~K, and 1000~K using the analytic expression for the dust
sublimation radius $R_{\mathrm{subl}}$ by \citet{mon02}
\begin{equation}
  R_{\mathrm{subl}}=0.034~\mathrm{AU} \left(\frac{1500~\mathrm{K}}{T_{\mathrm{subl}}}\right)^2 \sqrt{\frac{L_{\star}}{L_{\sun}} \frac{1}{\epsilon}}. \label{eqn:sublradiusIN05}
\end{equation}
In this relation, $\epsilon$ denotes the ratio of the absorption efficiencies
of the dust at the sublimation temperature $T_{\mathrm{subl}}$ and at the
stellar effective temperature $T_{\star}$, 
i.e.\ $\epsilon=\kappa_{P}(T_{\mathrm{subl}})/\kappa_{P}(T_{\star})$ with
$\kappa_{P}(T)=\int Q_{\mathrm{abs}}(\lambda) B_{\lambda}(T) d\lambda/\int B_{\lambda}(T) d\lambda$.
We assume $\epsilon = 1$, corresponding to rather large dust grains of several
$\mu$m in size and, therefore, to the innermost region of the inner dust rim.

We find that the measured continuum radii of three of the five objects (HD\,163296, 
HD\,104237, HD\,98922) follow the $R_{\mathrm{subl}} \propto L_{\star}^{1/2}$ law rather closely
assuming dust sublimation temperatures  $T_{\mathrm{eff}}$ between 1300~K and 1500~K, 
which is in agreement with the results obtained in earlier NIR broadband 
interferometric survey observations \citep{mon02}.
For the Herbig~Be stars MWC\,297 and V921\,Sco, the derived ring diameters 
are significantly more compact than the expected dust sublimation radii.
A possible explanation for the small apparent continuum sizes might be inclination effects,
although, based on complementary information from the literature (see Sect.~\ref{sec:interpretation}),
we expect notable inclination only for HD\,163296.
Alternatively, the apparent disk size might be reduced either due to gas absorption
(allowing dust to exist closer to the star) or due to the emission of optically thick gas
located inside of the dust-sublimation radius.
As discussed in Sects.~\ref{sec:mwc297} and \ref{sec:v921sco}, we favor these latter explanations 
for MWC\,297 and V921\,Sco.

\subsection{{\BrG} line emission}
\label{sec:discussbrg}

\subsubsection{Correlations with stellar parameters}

\citet{eis07b} speculated that the size of the {\BrG}-emitting region might
depend on $L_{\star}$, with the (spatially compact) accretion processes
dominating for stars with low luminosity and (spatially extended) outflow
processes dominating in the high-luminosity regime.
Therefore, we investigate whether the derived physical size 
($R_{\mathrm{Br}\gamma}$) or the relative size of the line-emitting region
($R_{\mathrm{Br}\gamma}/R_{\mathrm{cont}}$, i.e.\ normalized by the size of 
the continuum-emitting region) scales with the stellar luminosity $L_{\star}$.
As can be seen in Fig.~\ref{fig:RRatiovsLstar}, we do not find a clear correlation which
includes all stars in our sample, although there might be a trend 
$R_{\mathrm{Br}\gamma} \propto L_{\star}$, if we exclude the Herbig~Be star HD\,98922.
Likewise, there might be a relation between $R_{\mathrm{Br}\gamma}$
and the stellar temperature $T_{\star}$ (Fig.~\ref{fig:RRatiovsTeff}, {\it top}), 
with a larger size of the {\BrG}-emitting region for higher
effective stellar temperature.
The fact that we have to exclude HD\,98922 seems to indicate that {\BrG}
traces not one unique physical mechanism for all stars in our sample, but at least 
two line-emitting mechanisms.
As we argue in Sect.~\ref{sec:interpretation} and in the following subchapters, these 
mechanisms might include accretion and outflow processes,
where the latter seem to be responsible for the tentative correlations with the 
stellar parameters noted above. 
However, given the small statistical sample, more observations
will be required to confirm these first trends.

\subsubsection{Correlations with spectroscopic parameters}

Concerning the spectroscopic parameters, we search for correlations
between $R_{\mathrm{Br}\gamma}$ and $R_{\mathrm{Br}\gamma}/R_{\mathrm{cont}}$ 
and the {\BrG} line luminosity $L($\BrG$)$.
\citet{gar06} determined this quantity for all stars in our sample from
the {\BrG} line equivalent width of the circumstellar component.
In principle, the accretion luminosity and mass accretion rate can be
determined from $L($\BrG$)$ using the relation by \citet{cal04} 
($\log L_{\mathrm{acc}}/L_{\sun} = 0.9 \times \left( \log (L_{\mathrm{Br}\gamma}/L_{\sun}) + 4 \right)-0.7)$).
However, since applying this relation for the B-type stars in our sample
would require to extrapolate the empirical $L($\BrG$)$--$L_{\mathrm{acc}}$ 
relation well outside the range of stellar properties (luminosity and mass) 
of the calibration sample, we refrain from this step and use the line luminosity instead.
There seems to be a general trend, that with increasing line luminosity 
the size of the {\BrG}-emitting region increases (Fig.~\ref{fig:RRatiovsLBrG}).
In 2001, \citeauthor{muz01} already proposed a correlation between
$W($\BrG$)$ (and thus also $L($\BrG$)$) and the size of the {\BrG}-emitting region in the
context of their magnetospheric accretion model for T~Tauri stars (the
low-mass counterparts of HAeBe stars).  In order to explain the empirical
$W($\BrG$)$--$L_{\mathrm{acc}}$ correlation, they propose
non-axisymmetric accretion, where the emitting area of discrete funnel flows
increases with increasing accretion rate.  However, since these funnel flows
should still be located within a few stellar radii, we interpret the
measurement of rather extended {\BrG}-emitting regions in four out of six
HAeBe stars as evidence against the hypothesis that {\BrG} is predominantly 
tracing magnetospheric accretion in these stars. 
It is also interesting to note that the width of the {\BrG} line seems to decrease
with increasing stellar temperature (see Fig.~\ref{fig:spectra}).
\citet{ste01} have shown that this could be explained by optical depth 
effects in wind models, although the influence of inclination effects
cannot be ruled out at the current time.

Another spectroscopic parameter which was used extensively in the past to
characterize the accretion and outflow processes around YSOs is the line
profile of the Balmer lines, especially {\Ha}.
\citet{ack05} classified the {\Ha} line
profile of all stars in our sample using the scheme ``double-peaked'',
``single-peaked'', ``P-Cygni'', and ``inverse P-Cygni''.
The six stars in the considered sample exhibit three of these four {\Ha} line 
profiles (see Tab.~\ref{tab:targetstars}), with the inverse P-Cygni profile
being the exception\footnote{As discussed, for example, by \citet{wal72} and
  \citet{sor96}, the inverse P-Cygni profile is generally believed to trace
  mass infall, but is found towards only $\sim 5$\% of all HAeBe stars
  \citep{ack05}}.
As discussed in Sect.~\ref{sec:intro}, interpreting these line profiles
without spatial information can lead to highly ambiguous results (for
instance, \citealt{cid93} could already reproduce three of these four line
profiles by varying the velocity law for an expanding wind).
Thus, it seems promising to compare the spatially resolved information
about the size of the {\BrG}-emitting region with the shape of the {\Ha} line
profile, potentially revealing correlations between the spatial distribution
and the kinematics of the emitting hydrogen.
In this respect, it is important to notice that even under the assumption
that {\BrG} and {\Ha} trace the same physical process, these lines will
form in different spatial regions due to their different energy levels -- with
{\BrG} tracing higher gas densities located closer to the star. 

As shown in Fig.~\ref{fig:RRatiovsPHa}, we find indications that
$R_{\mathrm{Br}\gamma}/R_{\mathrm{cont}}$ depends on the {\Ha} 
line profile classification:

\noindent {\bf P-Cygni:}
For the HAeBe stars with a P-Cygni {\Ha}-line profile (MWC\,480, HD\,98922), the
{\BrG}-emitting region is very compact ($R_{\mathrm{Br}\gamma}/R_{\mathrm{cont}} \lesssim
0.2$, unresolved by the interferometer), consistent with an origin in a compact
stellar wind, an X-wind or in magnetospheric accretion.
P-Cygni {\Ha} line profiles are generally attributed to mass outflow 
\citep{fin84b,cat87}, although for T~Tauri stars \citet{muz01} showed that these
profiles can also be reproduced in magnetospheric accretion models assuming
blue-shifted absorption by gas in an accretion-powered wind located outside of
the magnetosphere.
Therefore, the {\BrG} line in MWC\,480 and HD\,98922 might primarily
trace magnetospheric accretion, while the optical Balmer lines are strongly
affected by absorption from outflowing gas.  
This interpretation is also consistent with the rather high accretion rates,
which have been reported  for these two objects (MWC\,480: $\log 
\dot{M}_{\mathrm{acc}}\geq-7.0~M_{\sun}$\,yr$^{-1}$, \citealt{eis07b};
HD\,98922: $\log \dot{M}_{\mathrm{acc}}=-5.76~M_{\sun}$\,yr$^{-1}$,
\citealt{gar06}).

\noindent {\bf Double-peaked:} Several theoretical studies \citep[e.g.~][]{cid93}
associate this {\Ha} line profile with rotating gas flows, including disk winds.
HD\,163296, HD\,104237, and V921\,Sco, which fit into this classification,
exhibit a {\BrG}-emitting region which is only slightly more compact than the
dust-sublimation radius (but still spatially resolved by our
spectro-interferometric measurements).
Considering the gaseous inner disk scenario, \citet{muz04} modeled the
continuum and line emission from such a disk and found that for low mass
accretion rates ($\log \dot{M}_{\mathrm{acc}} < -7~M_{\sun} $yr$^{-1}$), the flux 
contribution from the gaseous inner disk is negligible. This leads us
to favor the stellar wind or disk wind scenario for these three objects.

\noindent {\bf Single-peaked:}
Exhibiting a single-peaked {\Ha}-line profile, MWC\,297 shows a rather
extended {\BrG}-emitting region, exceeding even the size of the
continuum-emitting region.  However, comparing the size of the {\BrG}-emitting
region around MWC\,297 to sources with double-peaked {\Ha}-line profiles
(such as V921\,Sco), we do not find fundamental differences within the
measurement uncertainties (especially taking the indications for a
particularly compact continuum-emitting region around MWC\,297 into account,
as discussed in Sect.~\ref{sec:mwc297}).
Theoretical models have associated single-peaked {\Ha} line profiles with
stellar winds or with disk winds \citep{cid93}, especially for massive YSOs
\citep{sim05}.

Based on the currently available spectroscopic and long-baseline spectro-interferometric data,
it seems not feasible to discern between stellar wind and disk wind scenarios.
This is due to the fact that the gas velocity field in stellar winds as well as in disk winds, 
comprise both rotating and expanding velocity components, which can result in similar line profiles 
as well as in similar spatial scales for the line-emitting region.
To solve these ambiguities, future observations will require a very good sampling 
of the line visibility function in order to measure the subtle imprints resulting 
from the differences of the gas kinematics close to the star.

\subsubsection{Implications on the empirical $L($\BrG$)$--$L_{\mathrm{acc}}$ relation and the accretion-outflow connection}

Within the last decade, a correlation between the {\BrG} line luminosity and the mass accretion rate 
(as independently determined from the UV excess, which is commonly attributed to the accretion shock)
could be established for pre-main-sequence stars with masses between $\sim 0.01$ \citep{nat04}
and $\sim 4~M_{\sun}$ \citep{muz98a,van05}.
Being applicable even in regions of high extinction and for very low mass accretion rates, 
this correlation is of considerable practical importance, although neither the traced processes
nor the underlying physical mechanisms are yet identified.
It was suggested that the {\BrG} line emission might be a direct tracer of mass accretion, 
and that for large accretion rates ($\dot{M}_{\mathrm{acc}} > 10^{-7}~M_{\sun} $yr$^{-1}$) 
the contributions from other processes might be negligible \citep{van05}.
However, attempts to reproduce the $L($\BrG$)$--$L_{\mathrm{acc}}$ relation using
magnetospheric accretion models were not successful, but found instead that the
{\BrG} line luminosity depends mainly on other parameters such as the size of the magnetosphere and the
gas temperatures \citep{muz98b,muz98a}.

Based on our spatially resolved observations of the {\BrG}-emitting region around five Herbig~Ae/Be stars,
we cannot support the scenario that {\BrG} generally traces gas located in the magnetospheric infall zone.
Considering only those stars in our sample which are within the current calibration range of the relation 
(i.e.\ rejecting the early Herbig~Be stars), we find that for one star (HD\,98922) the size of the 
{\BrG}-emitting region is consistent with magnetospheric accretion, 
while two stars (HD\,104237, HD\,163296) exhibit a {\BrG}-emitting region which is more consistent with
an extended stellar wind or a disk wind scenario ($R_{\mathrm{Br}\gamma} \gtrsim 0.6~R_{\mathrm{cont}}$).
Therefore, our observations imply that, at least for some HAeBe stars, 
{\BrG} is not a {\it primary} tracer of accretion, but likely 
{\it indirectly} linked to the accretion rate, e.g.\ via accretion-driven mass loss.
This ambivalent origin of the line emission (tracing partially
infalling and outflowing matter) makes the empirical correlation between 
$L($\BrG$)$ and $L_{\mathrm{acc}}$ even more remarkable and suggests 
a tight, quantitative connection between the accretion and ejection processes in YSOs.
The existence of such a connection has already been suggested based on 
measurements of the luminosity of accretion and outflow-tracer lines,
yielding an ejection efficiency of $\dot{M}_{\mathrm{eject}}/\dot{M}_{\mathrm{acc}} \approx 0.1-0.2$
\citep{cab90,har94b}.

\subsubsection{Implications on the jet-launching mechanism for Herbig~Ae/Be stars}

Our measurement of rather extended {\BrG}-emitting regions also has important
implications on the launching mechanism for HAeBe jets.  
Although most stars in our sample show only indirect signatures of outflow activity
(e.g.\ CO outflows, outflow-tracing line emission), HD\,104237 and HD\,163296 
are also known to be associated with collimated microjets.
On large scales, the interaction region between the jet ejecta and
ambient material is traced by Herbig-Haro objects, while on arcsecond-scales,
the jets were also imaged with the HST in forbidden lines and in the hydrogen
{\Ha} and {\LyA} transitions \citep{gra04,was06}.

Therefore, it seems likely that our AMBER {\BrG} observations trace ionized gas located 
at the base of these microjets, just experiencing the acceleration and magnetic confinement
processes which are required to achieve the highly collimated gas flows detected by HST.
Since it is expected that stellar winds, lacking any collimation mechanism, result in wide-angle
outflows, we clearly favor the disk wind model for these objects in order to explain the observational evidence
for {\it (a)} a rather extended {\BrG}-emitting region on sub-AU scales and, simultaneously, {\it (b)}
collimated narrow-angle microjets on scales of several hundred AU.
This main conclusion is in qualitative agreement with HST spectroscopic observations, which measured the
jet opening angle and the toroidal velocity field on spatial scales $\sim 50$~AU from the base.
Traced back to the footpoint of the jet, the measured velocity fields seem more consistent
with a launching of the jet from the disk and not an X-wind or stellar wind \citep{bac02,tes02,cof07}.

Future spectro-interferometric observations will allow us to confirm this
hypothesis by measuring the spatial distribution and kinematics of the line-emitting
gas and comparing it with the orientation and velocity of the large-scale outflow
structure.

\section{Conclusions}
\label{sec:conclusions}

We summarize the results from our investigation using VLTI/AMBER data of five
HAeBe stars as follows:
\begin{enumerate}
\item The $K$-band continuum-emitting region was spatially resolved for all
  objects.
  For the Herbig~Ae and late Herbig~Be stars the characteristic size of this 
  region scales roughly with the square-root of the stellar luminosity, supporting 
  the scenario that for these sources, the continuum is dominated by thermal emission from the inner
  rim of the dust disk.
  For the early-type Herbig~Be stars MWC\,297 and V921\,Sco, on the other hand,
  the continuum-emitting region is significantly more compact than
  predicted by the $R_{\mathrm{subl}} \propto L_{\star}^{1/2}$ relation, 
  maybe suggesting that for these objects the NIR continuum is dominated by hot gas emission.
  Fits of inclined ring geometries to the visibilities measured on HD\,163296 
  indicate object elongation, consistent with a notably inclined disk geometry, 
  while for HD\,98922 and V921\,Sco we find no indications for a large disk inclination.
\item Closure phases were measured for four objects in the $K$-band continuum 
  regime.  For V921\,Sco we measured the CP also within the spectrally resolved
  {\BrG} line, suggesting a nearly centro-symmetric line-emitting region 
  (in agreement with the low inclination angle).
\item In the spectrally resolved {\BrG}~lines of HD\,163296, HD\,104237,
  HD\,98922, and V921\,Sco, we measure an increase of visibility, indicating
  that the line-emitting region is more compact than the continuum-emitting
  region. MWC\,297 shows a decrease of visibility within the {\BrG}-line,
  although this is likely related to the compactness of the continuum-emitting
  region and not to fundamental differences in the line formation
  mechanism.
\item Judging on the currently available limited statistics, we
  do not find a trend which relates the size of the {\BrG} line-emitting
  region with the stellar luminosity, as was proposed by \citet{eis07b}.
  Instead, the continuum-normalized size of the line-emitting region
  $R_{\mathrm{Br}\gamma}/R_{\mathrm{cont}}$ might correlate with the stellar 
  effective temperature, although more observations will be required to confirm this trend. 
  Maybe the most remarkable correlation we found was that stars with a P-Cygni {\Ha}~line 
  profile and a high mass-accretion rate seem to show particularly compact {\BrG}-emitting
  regions ($R_{\mathrm{Br}\gamma}/R_{\mathrm{cont}} <0.2$), while stars with
  a double-peaked or single-peaked {\Ha}-line profile show a
  significantly more extended {\BrG}-emitting region
  ($0.6 \lesssim R_{\mathrm{Br}\gamma}/R_{\mathrm{cont}} \lesssim 1.4$). 
\item  Only for HD\,98922, the emitting-region is compact enough to support
  the hypothesis that most of the {\BrG}-emission emerges from magnetospheric
  accretion columns.  In the other cases, the {\BrG}-emitting region is only
  slightly more compact than the dust sublimation radius, supporting the idea
  that the emission mainly emerges from an extended stellar wind or a
  disk-wind.  This makes it unlikely that {\BrG} primarily traces
  magnetospheric accretion in HAeBe stars, suggesting that even for
  Herbig Ae stars, this line is only an indirect tracer of the mass 
  accretion rate.
\end{enumerate}

\begin{acknowledgements}
We are grateful to A.~Chelli and G.~Duvert for carrying out independent 
checks on the AMBER data reduction.
Furthermore, we would like to thank the referee for his
comments, which helped to improve the paper.
This publication makes use of data products from the Two Micron All Sky
Survey, which is a joint project of the University of Massachusetts and the
Infrared Processing and Analysis Center/California Institute of Technology,
funded by the National Aeronautics and Space Administration and the National
Science Foundation.
NSO/Kitt Peak FTS data used here were produced by NSF/NOAO.
\end{acknowledgements}

\bibliographystyle{aa}
\bibliography{9946}

\end{document}